\documentclass{article}
\usepackage{graphicx}
\usepackage{hyperref}
\usepackage{lineno}
\usepackage[OT1]{fontenc}
\usepackage{amsthm,amsmath,amssymb}
\usepackage[square,numbers]{natbib}
\usepackage{graphicx}
\usepackage{subfig}
\usepackage{titlesec}
\usepackage{mathrsfs}
\usepackage{multirow}
\usepackage{booktabs}
\usepackage{color}
\usepackage{rotating}

\RequirePackage{enumitem} 
\RequirePackage{mathtools}
\usepackage[doublespacing]{setspace}
\usepackage{geometry}
\numberwithin{equation}{section}
\theoremstyle{plain}

\titlelabel{\thetitle.\quad}

\makeatletter

\renewcommand\newblock{\hskip .11em\@plus.33em\@minus.07em}
\makeatother

\newenvironment{bottompar}{\par\vspace*{\fill}}{\clearpage}
\begin{document}

\title{\textbf{Admissible multi-arm stepped-wedge cluster randomized trial designs}}
\author{\textbf{M. J. Grayling\textsuperscript{1}, A. P. Mander\textsuperscript{1}, J. M. S. Wason\textsuperscript{1,2}}\\
	\small 1. Hub for Trials Methodology Research, MRC Biostatistics Unit, Cambridge, UK, \\ \small 2. Institute of Health and Society, Newcastle University, Newcastle, UK.}
\date{}
\maketitle

\noindent \textbf{Abstract:} Numerous publications have now addressed the principles of designing, analyzing, and reporting the results of, stepped-wedge cluster randomized trials. In contrast, there is little research available pertaining to the design and analysis of multi-arm stepped-wedge cluster randomized trials, utilized to evaluate the effectiveness of multiple experimental interventions. In this paper, we address this by explaining how the required sample size in these multi-arm trials can be ascertained when data are to be analyzed using a linear mixed model. We then go on to describe how the design of such trials can be optimized to balance between minimizing the cost of the trial, and minimizing some function of the covariance matrix of the treatment effect estimates. Using a recently commenced trial that will evaluate the effectiveness of sensor monitoring in an occupational therapy rehabilitation program for older persons after hip fracture as an example, we demonstrate that our designs could reduce the number of observations required for a fixed power level by up to 58\%. Consequently, when logistical constraints permit the utilization of any one of a range of possible multi-arm stepped-wedge cluster randomized trial designs, researchers should consider employing our approach to optimize their trials efficiency.\\

\noindent \textbf{Keywords:} Admissible design, Cluster randomized trial, Multiple comparisons, Optimal design, Stepped-wedge.\\

\begin{bottompar}
	\noindent Address correspondence to M. J. Grayling, MRC Biostatistics Unit, Forvie Site, Robinson Way, Cambridge CB2 0SR, UK; Fax: +44-(0)1223-330365; E-mail: mjg211@cam.ac.uk. 
\end{bottompar}

\section{Introduction}

In a cluster randomized trial (CRT), groups of participants, not individuals, are randomized. The advantages this can bring are today recognized as numerous. For example, CRTs can aid the control of contamination between
participants, and can bring increased administrative efficiency, helping to overcome the barriers of recruiting large numbers of participants.\cite{vickers2014} Unfortunately, there are also several well-noted disadvantages to CRTs.\cite{donner2004,edwards1999} Specifically, double blinding should ideally be present in every trial, however, it is often impossible in CRTs. Moreover, missing data can quickly become a problem if whole clusters are lost to follow-up.

Nevertheless, there has now been much work conducted on design and analysis procedures for CRTs. One type of CRT that has received considerable attention recently, and which we focus on here, is the stepped-wedge (SW)-CRT (see, e.g., Hussey and Hughes (2007)\cite{hussey2007}). In a SW-CRT, an intervention is introduced over several time periods, and typically all clusters receive the intervention by the end of the trial. Numerous potential advantages to this design have been forwarded. Principally, all clusters receiving the intervention is advantageous if it is expected to do more good than harm. The design's sequential implementation can also increase feasibility when there are logistical or practical constraints. However, these alleged advantages have been disputed. Primarily, it has been argued that an intervention should not be implemented in every cluster when it has not yet been proven to be effective. For brevity, we refer the reader elsewhere for further discussion of these points.\cite{mdege2012,hemming2013,keriel2014,kotz2012,kotz2012b,kotz2013,dehoop2015,hargreaves2015,prost2015}

Methodological developments in this area include Hussey and Hughes (2007),\cite{hussey2007} who provided guidance on sample size calculations for cross-sectional SW-CRTs analyzed with a particular linear mixed model. Here, cross-sectional designs refer to a scenario in which measurements are accrued on different participants in each time period. This work was later built upon to establish a design effect for cross-sectional SW-CRTs,\cite{woertman2013} and also to allow for transition periods and multiple levels of clustering.\cite{hemming2015} Recently, similar results for cohort SW-CRTs, in which repeated measurements are accrued on a single group of patients, have been presented.\cite{hooper2016} Finally, explanations on determining the sample size required by SW-CRTs through simulation have also been presented\cite{baio2015}.

Thus, sample size determination for SW-CRTs has been well studied. However, the above articles only discuss sample size calculations for a particular design. That is, a design with prescribed rules about how the experimental intervention will be allocated across the clusters. Moreover, with the exception of Baio et al. (2015),\cite{baio2015} each paper deals only with a specific analysis model. Addressing these limitations, recent research has ascertained optimal treatment allocation rules for several general classes of cross-sectional SW-CRT design, analyzed with a highly flexible linear mixed model.\cite{lawrie2015,girling2016,thompson2017} A subset of these results has subsequently been extended to cohort SW-CRTs.\cite{li2018} Nonetheless, there is still a need for guidance on the optimal design of SW-CRTs with more specialized analysis models.

Furthermore, the above publications relate only to the design of two-arm SW-CRTs. Very little research has been conducted on the design of CRTs with multiple experimental treatment arms, and in particular scenarios in which clusters may switch between interventions. We refer to such designs in this article as multi-arm stepped-wedge cluster randomized trials (MA-SWs). Formulae for the variance of the treatment effect estimators of several possible designs with three treatment arms, using a specific linear mixed model for data analysis, are available.\cite{teerenstra2015} An additional paper recently proposed, and compared the efficiencies of, several simple variants of the classical SW-CRT design that could be used to accommodate multiple interventions.\cite{lyons2017} Finally, utilizing experimental design theory, the performance of several analysis models for the same such MA-SW designs was recently examined.\cite{matthews2017} However, these are the only works that we are aware of pertaining to the design of MA-SWs. This is perhaps surprising since several studies have recently been conducted in such a manner.\cite{chinbuah2012,pol2017} Furthermore, intuitively these designs could have numerous advantages that it would be beneficial to highlight. Explicitly, evaluating multiple interventions within the same CRT could bring the same sort of efficiency gains multi-arm trials bring to individually randomized studies.\cite{parmar2014} That is, the required number of clusters or observations could be reduced relative to conducting several separate trials. Moreover, it could allow for a reduction in required funding as a consequence of reduced administrative costs, and may allow for the assessment of intervention interactions. Furthermore, one would anticipate that such designs could on average decrease the time taken for each cluster to receive a particular intervention, which may improve cluster and patient participation. However, the potential of MA-SWs can only be realised if we design such studies effectively; poorly designed MA-SW trials would likely result in a poor answer being acquired to numerous important questions.

Therefore, here, we first discuss how one can compute the sample size required by, and optimized treatment sequence allocations for, a MA-SW design when a linear mixed model is used for data analysis. We then consider one particular analysis model, and utilizing a recently undertaken trial as our principal motivation, discuss how large the efficiency gains made using our methods could be in practice. 

\section{Methods}

\subsection{Notation, hypotheses, and analysis}

We designate a MA-SW as any trial conforming to the following requirements

\begin{itemize}
	\item The trial is carried out in $C \ge 2$ clusters, over $T \ge 2$ time periods, with $m > 1$ measurements made in each cluster in each time period;
	\item In each time period, each cluster receives a combination of a set of $D$ interventions (indexed by $d=0,\dots,D-1$);
	\item The sequence of intervention allocations for each cluster is specified randomly.
\end{itemize}

We make no assumptions about whether the $m$ measurements from each time period are on different patients; a cross-sectional design, or the same patients; a cohort design. We do not require each cluster to begin on, receive, or conclude the trial on any particular intervention. We also do not enforce the usual one-directional switching associated with conventional SW-CRTs, so as to allow for transitions between experimental interventions in any order, if this is desired. As a consequence of this, the methodology we describe is applicable to the design of multi-arm cluster randomized crossover trials. We keep in mind, however, that each of the interventions must be received by at least one cluster in some time period for its effect to be estimable.

Throughout we assume that the accrued data from the trial will be normally distributed, and an identifiable linear mixed model will be utilized for data analysis, denoted as

\[ \boldsymbol{y} = A\boldsymbol{\beta} + Z\boldsymbol{u} + \boldsymbol{\epsilon}, \]

where

\begin{itemize}
	\item $\boldsymbol{y}$ is the vector of responses;
	\item $\boldsymbol{\beta}=(\beta_1,\dots,\beta_p)^\top$ is a vector of $p$ fixed effects;
	\item $A$ is the design matrix which links $\boldsymbol{y}$ to $\boldsymbol{\beta}$;
	\item $\boldsymbol{u}$ is a vector of random effects, with $\boldsymbol{u} \sim N(\boldsymbol{0},G)$, where $G$ is a specified (assumed known) matrix;
	\item $Z$ is the design matrix which links $\boldsymbol{y}$ to $\boldsymbol{u}$;
	\item $\boldsymbol{\epsilon}$ is a vector of residuals, with $\boldsymbol{\epsilon} \sim N(\boldsymbol{0},R)$, where $R$ is a specified (assumed known) matrix.
\end{itemize}

We suppose that $\boldsymbol{\beta}$ has been specified such that its first $q$, $q\le p$, elements, $(\beta_1,\dots,\beta_{q})^\top$, are our parameters of interest. Typically, we may have that $q=D-1$, with these parameters representing either the direct effects of a set of experimental interventions relative to some control, or the direct effect of intervention arm $d$ relative to intervention arm $d-1$, for $d = 1,\dots,D-1$. However, we do not require that this be the case. Then, we assume that we will test the following one-sided hypotheses
	
	\[ H_{0f} : \beta_f \le 0, \qquad H_{1f} : \beta_f>0, \qquad f=1,\dots,q. \]

We note though that the determination of MA-SW designs for alternative hypotheses of interest, e.g., two-sided hypotheses, is also easily achievable by adapting what follows.

To test these hypotheses, following trial completion, we estimate $\boldsymbol{\beta}$ using the
maximum likelihood estimator of a linear mixed model

\[ \hat{\boldsymbol{\beta}} = (\hat{\beta}_1,\dots,\hat{\beta}_p)^\top = \{A(ZGZ^\top + R)^{-1}A\}^{-1}A^\top(ZGZ^\top +R)^{-1}\boldsymbol{y}. \]

Then

\[ \text{cov}(\hat{\boldsymbol{\beta}},\hat{\boldsymbol{\beta}}) = \Lambda = \{A(ZGZ^\top + R)^{-1}A\}^{-1}. \]

We set $\hat{\boldsymbol{\beta}}_q = (\hat{\beta}_1,\dots,\hat{\beta}_q)^\top$, and denote the covariance matrix of $\hat{\boldsymbol{\beta}}_q$ by $\Lambda_q$. That is, $\text{cov}(\hat{\boldsymbol{\beta}}_q,\hat{\boldsymbol{\beta}}_q) = \Lambda_q$.

Our conclusions are then based upon the following Wald test statistics

\[ Z_f = \frac{\hat{\beta}_f}{\sqrt{\text{var}(\hat{\beta}_f)}} = \hat{\beta}_f\Lambda^{-1/2}_{[f,f]} = \hat{\beta}_f I_f^{1/2}, \qquad f=1,\dots,q. \]

Explicitly, we reject $H_{0f}$ if $Z_f>e$, for critical boundary $e$. Given $e$, we can determine for any vector of true fixed effects $\boldsymbol{\beta}_q$ the probability each particular $H_{0f}$ is rejected, and the probability we reject at least one of $H_{01},\dots,H_{0q}$, via the following integrals

\begin{align*}
\mathbb{P}(\text{Reject } H_{0f} \mid \beta_f) &= \int_e^{\infty}\phi(x,\beta_fI_f^{1/2},1)\mathrm{d}x,\\
\mathbb{P}(\text{Reject at least one of } H_{01},\dots,H_{0q} \mid \boldsymbol{\beta}_q) &= 1 - \int_{-\infty}^e\dots\int_{-\infty}^e\phi\{\boldsymbol{x},\boldsymbol{\beta}_q\circ \boldsymbol{I}^{1/2},\text{diag}(\boldsymbol{I}^{1/2})\Lambda_q\text{diag}(\boldsymbol{I}^{1/2})\}\mathrm{d}x_{q}\dots\mathrm{d}x_1.
\end{align*}

Here

\begin{itemize}
	\item $\phi(\boldsymbol{x},\boldsymbol{\mu},\Sigma)$ is the probability density function of a multivariate normal distribution with mean $\boldsymbol{\mu}=(\mu_1,\dots,\mu_k)^\top$ and covariance matrix $\Sigma$, $\text{dim}(\Sigma)=k \times k$, evaluated at vector $\boldsymbol{x} = (x_1,\dots,x_k)^\top$;
	\item $(a_1,\dots,a_n)^\top\circ(b_1,\dots,b_n)^\top = (a_1b_1,\dots,a_nb_n)^\top$;
	\item $\boldsymbol{I}^{1/2}=(I_1^{1/2},\dots,I_{q}^{1/2})^\top$ is the element-wise square root of the vector of information levels for $\boldsymbol{\beta}_q$;
	\item $\text{diag}(\boldsymbol{v})$ for a vector $\boldsymbol{v}$ indicates the matrix formed by placing the elements of $\boldsymbol{v}$ along the leading diagonal.
\end{itemize}

Determining an appropriate value for $e$ depends upon whether a correction for multiple testing is to be utilized. Without such a correction, $e$ can be chosen to control the per-hypothesis error-rate to $\alpha$ by setting $e$ as the solution to

\[ \alpha = \int_e^\infty \phi(x,0,1)\mathrm{d}x. \]

Alternatively, the familywise error-rate, the probability of one or more false rejections, can be controlled for example using the Bonferroni correction, which sets in this instance $e$ to be the solution of

\[ \frac{\alpha}{q} = \int_e^\infty \phi(x,0,1)\mathrm{d}x. \]

The choice of whether to utilize a multiple testing correction is not a simple one, with much debate in the literature around when it is necessary. It seems reasonable for MA-SWs however to extrapolate from previous discussions, and note that one should correct in confirmatory settings, but should not always feel the need to in exploratory settings.\cite{wason2014}

\subsection{Power considerations}

The above fully specifies a hypothesis testing procedure for a MA-SW. However, at the design stage, it is important to be able to determine values of $m$, $C$, and $T$ that provide both the desired per-hypothesis or familywise error-rate, and the desired power. Here, we describe two types of power that could be required, since power is not a simple concept in multi-arm trials.

We suppose that power of at least $1-\beta$ is required either to reject each $H_{0f}$ (individual power), or at least one of $H_{01},\dots,H_{0q}$ (combined power), when $\boldsymbol{\beta}_q=\boldsymbol{\delta}=(\delta_1,\dots,\delta_{q})^\top$. The element $\delta_f$ here represents a clinically relevant difference for the effect $\beta_f$. Using our notation from earlier, these requirements can be written as

\begin{align*}
\min_{f\in\{1,\dots,q\}}\mathbb{P}(\text{Reject } H_{0f} \mid \delta_f) &\ge 1 - \beta,\\
\mathbb{P}(\text{Reject at least one of } H_{01},\dots,H_{0q} \mid \boldsymbol{\delta}) &\ge 1 - \beta.
\end{align*}

The choice between these requirements should be made based on several considerations. The latter will likely require smaller sample sizes, however it would leave a trial less likely to reject all false null hypotheses. Therefore, trialists must weigh up the cost restrictions and goals of their trial. 

\subsection{Design specification}

We can now return to our considerations on determining appropriate values for $m$, $C$, and $T$. We must also determine as part of the same process a matrix $X$ that indicates the planned allocation of interventions to each cluster across the time periods. Extending the notation commonly utilised for SW-CRTs,
$X$ is a $C\times T$ matrix, with $X_{ij}$ indicating which intervention(s) cluster $i$ receives in time period $j$. If only a single intervention is given to each cluster in each time period, then $X_{ij}$ will be a single number. Otherwise, it may be some combination of values, indicating allocation to multiple interventions. With this, it will now be useful to denote the design utilized by a trial by $\mathscr{D}=\{m,C,T,X\}$, and the associated covariance matrix for $\boldsymbol{\beta}_q$ by $\Lambda_\mathscr{D}$. Our goal is then to optimize $\mathscr{D}$.

Most of the work on sample size determination for SW-CRTs pre-supposes that two of the three parameters $m$, $C$, and $T$ are fixed (with one usually $T$), and then looks to identify the third. In addition, the matrix $X$ is usually specified, if not explicitly (in the case where $C$ and $T$ are fixed), then through some rule such as balanced stepping. Here, we take an alternate approach to the determination of the preferred design. We assume that a set of allowed values for $T$ has been specified, $\mathfrak{T}=\{T_1,\dots,T_{|\mathfrak{T}|}\}$. We then suppose that sets of allowed values for $C$, for each element of $\mathfrak{T}$, have been specified. We denote these by $\mathfrak{C} = \{\mathfrak{C}_{T_1},\dots,\mathfrak{C}_{T_{|\mathfrak{T}|}}\}$, with $\mathfrak{C}_{T_i} = \{C_1,\dots,C_{|\mathfrak{C}_{T_i}|}\}$. Furthermore, we suppose that for each allowed $C$,$T$ combination, a set of allowed values for $m$ have been provided; $\mathfrak{M}_{C,T}=\{m_1,\dots,m_{|\mathfrak{M}_{C,T}|}\}$. We then take $\mathfrak{M}=\{\mathfrak{M}_{C,T} : T\in\mathfrak{T},C\in\mathfrak{C}_T\}$. We allow for such an interrelated specification of the values for $m$, $C$, and $T$ to cover many possible design scenarios. For example, increasing the value of $T$ may mean logistical constraints force only lower values of $m$ and $C$ to be possible. In actuality, it is likely a trialist would not need such a complicated structure. For example, the classical case of fixed $T$ and $m$, searching for the correct value for $C$, would require only $\mathfrak{T}=\{T\}$, $\mathfrak{M}_{C,T}=\{m\}$, and $\mathfrak{C}=\{\mathfrak{C}_T\}$, with $\mathfrak{C}_T=\{2,\dots,C_{\text{max}}\}$, and $C_{\text{max}}$ some suitably large value.

Finally, for each $C,T$ combination, we also specify a set of allowed $X$, which we denote by $\mathfrak{X}_{C,T}$. Similar to the above, we then take $\mathfrak{X}=\{\mathfrak{X}_{C,T} : T\in\mathfrak{T},C\in\mathfrak{C}_T\}$. Shortly, we will describe several possible ways in which $\mathfrak{X}_{C,T}$ could be specified.

Now, with $\mathfrak{T}$, $\mathfrak{C}$, $\mathfrak{M}$, and $\mathfrak{X}$ chosen, formally our set $\mathfrak{D}$ of all allowed possible designs is

\[ \mathfrak{D} = \{ \mathscr{D} : T\in\mathfrak{T}, C\in \mathfrak{C}_T, m\in \mathfrak{M}_{C,T}, X \in \mathfrak{X}_{C,T} \}.\]

\subsection{Admissible design determination}

As was discussed, previous research has assessed which is the optimal SW-CRT design to maximize power in an array of possible design scenarios. This was achieved by developing formulae for the efficiency of designs under particular linear mixed models. Such considerations could in theory be extended to MA-SWs, or to alternate analysis models. However, it is not practical to conduct such derivations for every value of $D$, or every analysis model that may need to be utilized. In addition, it is not actually necessary following specification of the set $\mathfrak{D}$: preferable designs can be determined using exhaustive or stochastic heuristic searches.

Explicitly, for some $\mathfrak{D}$, modern computing makes an exhaustive search possible using parallelisation. Alternatively, in the case where $C$ and $T$ are fixed (either in advance or after some initial design identification), we can employ a different method to determine our final design: a stochastic search. This is sensible when, even with $C$ and $T$ fixed, the design space $\mathfrak{D}$ remains large. Here, we accomplish this optimization using CEoptim in R.\cite{benham2017}

To perform a search, an optimality criterion is required. Previous research on SW-CRTs has focused on determining designs that minimize the variance of the treatment effect estimator. Here, we extend this to consider designs that minimize some weighted combination of a trial cost function, and some factor formed from the covariance matrix of the treatment effect estimators, $\Lambda_\mathscr{D}$.

Specifically, we allocate a function $f(\mathscr{D})$ that sets the cost associated with a trial using design $\mathscr{D}$. This could be as simple as the required number of observations, or something more complex that factors in the speed the interventions would need to be rolled out according to $X$, for example.

For $\Lambda_\mathscr{D}$, numerous possible optimality criteria have been suggested in the literature. We consider D-, A-, and E-optimal designs, which all have a long history within the field of experimental design. D-optimality corresponds to minimizing the \textit{determinant} of $\Lambda_\mathscr{D}$, $\text{det}(\Lambda_\mathscr{D})$. This can be interpreted as minimizing the volume of the confidence ellipsoid for the $\beta_f$. For A-optimality the average value of the elements along the diagonal of $\Lambda_\mathscr{D}$, $\text{tr}(\Lambda_\mathscr{D})/q$, is minimized. That is, we minimize the \textit{average} variance of the $\beta_f$. And finally, in E-optimality, we minimize the maximal value of the elements along the diagonal of $\Lambda_\mathscr{D}$, $\max Diag(\Lambda_\mathscr{D})$, i.e., we minimize the most \textit{extreme}, or largest, of the variances of the $\beta_f$. We refer the reader elsewhere for greater detail on these criteria.\cite{atkinson1992,dmitrienko2007}

Then, for example, our admissible design using the D-optimality criteria will be the $\mathscr{D}_*$, conforming to the trials power requirements, that minimizes

\begin{equation}\label{eq1}
w \frac{f(\mathscr{D}_*) - \min_{\mathscr{D}\in\mathfrak{D}}f(\mathscr{D})}{\max_{\mathscr{D}\in\mathfrak{D}}f(\mathscr{D}) - \min_{\mathscr{D}\in\mathfrak{D}}f(\mathscr{D})} + (1 - w) \frac{\text{det}(\Lambda_{\mathscr{D}_*}) - \min_{\mathscr{D}\in\mathfrak{D}}\text{det}(\Lambda_{\mathscr{D}})}{\max_{\mathscr{D}\in\mathfrak{D}}\text{det}(\Lambda_{\mathscr{D}}) - \min_{\mathscr{D}\in\mathfrak{D}}\text{det}(\Lambda_{\mathscr{D}})}.
\end{equation}

Here, $f(\mathscr{D}_*)$ and $\text{det}(\Lambda_{\mathscr{D}_*})$ are rescaled precisely because they exist on different scales. Additionally, $0 \le w \le 1$ is the weight given to
minimizing the trials cost relative to the efficiency of $\Lambda_\mathscr{D}$. Note that the case
$w=1$ should often be ignored since many designs will likely share equal values of $f(\mathscr{D})$. Admissible designs using the A- or E-optimality criteria are formed by replacing $\text{det}(\Lambda_\mathscr{D})$ in the above by $\text{tr}(\Lambda_\mathscr{D})/q$ or $\max Diag(\Lambda_\mathscr{D})$ respectively.

Note that if all of the designs in $\mathfrak{D}$ cannot attain the desired power, no admissible design will exist. To counteract this, we can increase the value of $\beta$. In an extreme scenario where no design will likely meet any reasonable power requirement, we can set $\beta=1$ and $w=0$ and look to determine the design $\mathscr{D}$ that simply minimises some function of $\Lambda_\mathscr{D}$.

Finally, the rescaling in Equation~\ref{eq1} is only possible in the case of an exhaustive search where minimal and maximal values can be identified. Therefore, in the case of a stochastic search, we consider only meeting the conventional D-, A- and E-optimality criteria, without rescaling.

\subsection{Example trial design scenarios and associated linear mixed model}\label{example}

In what follows, we frame our examples within the context of studies in which there is a nested natural order upon the $D$ interventions. That is, as in Chinbuah \textit{et al.} (2012)\citep{chinbuah2012} and Pol \textit{et al.} (2017),\citep{pol2017} for $d=1,\dots,D-1$, intervention $d$ consists of intervention $d-1$ and some additional factor (e.g., intervention $d$ may include additional components of some wider multi-faceted intervention over intervention $d-1$). We therefore now in all instances enforce the restrictions that each cluster receives only a single intervention in each time period, and that if a cluster receives intervention $d$ in time period $j$, it cannot receive interventions $0,\dots,d-1$ in time periods $j+1,\dots,T$. Relating this restriction to our matrix $X$, it implies $X_{ij}\ge X_{ij-1}$ for $j=2,\dots,T$ and $i=1\dots,C$.

Our methodology for the determination of admissible MA-SW designs is now fully specified. Code to implement our methods and replicate our results is available from https://github.com/mjg211/article\_code. Next, several example trial design scenarios are considered to demonstrate the efficiency gains our designs could bring. In each we assume that the goal is to compare the efficacy of intervention $1$ to intervention $0$, intervention $2$ to intervention $1$, and so on, giving $q=D-1$. Moreover, in all examples the following linear mixed model, an extension of that used in Girling and Hemming (2016)\cite{girling2016} and Hooper \textit{et al.} (2016)\cite{hooper2016} to a multi-arm setting, is employed for data analysis
	
	\[ y_{ijk} = \mu + \pi_j + \beta_{1}\mathbb{I}\{X_{ij}\ge 1\} + \dots + \beta_{D-1}\mathbb{I}\{X_{ij}\ge D-1\} + c_i + \theta_{ij} + s_{ik} + \epsilon_{ijk}. \]

Here

\begin{itemize}
	\item $\mathbb{I}(x)$ is the indicator function on event $x$;
	\item $y_{ijk}$ is the $k$th response ($k=1,\dots,m$), in the $i$th cluster ($i=1,\dots,C$), in the $j$th time period ($j=1,\dots,T$);
	\item $\mu$ is an intercept term;
	\item $\pi_j$ is the fixed effect for the $j$th time period (with $\pi_1=0$ for identifiability);
	\item $c_i$ is the random effect for cluster $i$, with $c_i \sim N(0,\sigma_c^2)$;
	\item $\theta_{ij}$ is a random interaction effect for cluster $i$ and period $j$, with $\theta_{ij} \sim N(0,\sigma_\theta^2)$;
		\item $s_{ik}$ is a random effect for repeated measures in individual $k$ from cluster $i$, with $s_{ik} \sim N(0,\sigma_s^2)$;
	\item $\epsilon_{ijk}$ is the residual error, with $\epsilon_{ijk} \sim N(0,\sigma_\epsilon^2)$;
\end{itemize}

Thus, we specify our model to be applicable to a cohort MA-SW trial. We can then recover a model appropriate for a cross-sectional design by setting $\sigma_s^2=0$. Note that by the above, the variance of response $y_{ijk}$ is $\sigma^2=\sigma_c^2+\sigma_\theta^2+\sigma_s^2+\sigma_{\epsilon}^2$. In Section~\ref{results}, we will make reference to the following three correlation parameters
	
	\begin{itemize}
		\item $\rho_0=(\sigma_c^2+\sigma_\theta^2)/\sigma^2$: the within-period correlation (the correlation between the responses from two distinct individuals, in the same cluster, in the same time period);
		\item $\rho_1=\sigma_c^2/\sigma^2$: the inter-period correlation (the correlation between the responses from two distinct individuals, in the same cluster, in distinct time periods);
		\item $\rho_2=(\sigma_c^2+\sigma_s^2)/\sigma^2$: the individual auto-correlation (the correlation between the responses from the same individual in distinct time periods).
	\end{itemize}
	
	Finally, note that we also restrict the sets $\mathfrak{X}_{C,T}$ in all instances to those $X$ which imply the above model is identifiable, which can be verified for any $X$ using the implied design matrix $A$. However, for brevity, we do not explicitly state this requirement in our forthcoming specifications of the sets $\mathfrak{X}_{C,T}$.

\section{Results}\label{results}

	\subsection{$D=2$: Girling and Hemming (2016) and Thompson \textit{et al.} (2017)}

It was previously demonstrated that the efficiency of a conventional SW-CRT (i.e., the case $D=2$), analysed with the above linear mixed model, could be assessed using the cluster mean correlation, given by\cite{girling2016}
	
	\[ E(\rho) = \frac{mT\rho}{1 + (mT - 1)\rho}, \]
	
	where $\rho$ is the intra-cluster correlation for the means of the observations at each time-point, in each cluster. The optimal $X$ matrices to minimise the variance of $\hat{\beta}_1$, when $T=6$ and $C=10$, were also provided in this paper. We now demonstrate how our exhaustive search procedure can identify such optimal designs.

First, we set $\mathfrak{I}=\{6\}$ and $\mathfrak{C}=\{\mathfrak{C}_6\}=\{10\}$. We place no further restrictions on $\mathfrak{X}_{10,6}$ than those outlined in Section~\ref{example}, and thus
	
	\[ \mathfrak{X}_{6,10} = \{ X : \text{dim}(X)=6\times10, X_{ij} \ge X_{ij-1} \text{ for } j=2,\dots,6 \text{ and } i=1\dots,10\}. \]

To minimize $\text{var}(\hat{\beta}_1)$, we take $w=0$ and $\beta=1$. Since $D=2$, the D-, A-, and E-optimality criteria are equivalent, and we do not need to specify a multiple comparison correction. Whilst with $\beta=1$, our choices for $\alpha$, $\boldsymbol{\delta}$, and desire for individual or combined power are irrelevant. Finally, for simplicity, we reduce our model to that from Hussey and Hughes (2007)\cite{hussey2007} by supposing that $\sigma_\theta^2=\sigma_s^2=0$. Then, $\rho=\rho_0=\rho_1=\rho_2$ is the conventional intra-cluster correlation associated with cross-sectional SW-CRTs. Accordingly, to find optimal designs for different ranges of $E(\rho)$, as in Girling and Hemming (2016),\cite{girling2016} we take as an example $\sigma^2=1$, $\mathfrak{M}_{6,10}=\{10\}$, and set $\rho$ as those values which imply $E(\rho)\in\{0.1, 0.15, 0.3, 0.45, 0.75, 0.9\}$.

\begin{table}[htbp]
	\renewcommand{\arraystretch}{1.2}
	\begin{center}
		\begin{tabular}{rrrrr}
			\toprule
			Factor && \multicolumn{3}{c}{Results} \\
			\cline{1-1}\cline{3-5}
			$E(\rho)$ && 0.1 & 0.15 & 0.3 \\
			$X$ & & $\begin{pmatrix} 0 & 0 & 0 & 0 & 0 & 0 \\
			0 & 0 & 0 & 0 & 0 & 0 \\
			0 & 0 & 0 & 0 & 0 & 0 \\
			0 & 0 & 0 & 0 & 0 & 0 \\
			0 & 0 & 0 & 0 & 0 & 0 \\
			1 & 1 & 1 & 1 & 1 & 1 \\
			1 & 1 & 1 & 1 & 1 & 1 \\
			1 & 1 & 1 & 1 & 1 & 1 \\
			1 & 1 & 1 & 1 & 1 & 1 \\
			1 & 1 & 1 & 1 & 1 & 1 \\ \end{pmatrix}$ &
			$\begin{pmatrix} 0 & 0 & 0 & 0 & 0 & 0 \\
			0 & 0 & 0 & 0 & 0 & 0 \\
			0 & 0 & 0 & 0 & 0 & 0 \\
			0 & 0 & 0 & 0 & 0 & 0 \\
			0 & 0 & 0 & 0 & 0 & 1 \\
			0 & 1 & 1 & 1 & 1 & 1 \\
			1 & 1 & 1 & 1 & 1 & 1 \\
			1 & 1 & 1 & 1 & 1 & 1 \\
			1 & 1 & 1 & 1 & 1 & 1 \\
			1 & 1 & 1 & 1 & 1 & 1 \\ \end{pmatrix}$ &
			$\begin{pmatrix} 0 & 0 & 0 & 0 & 0 & 0 \\
			0 & 0 & 0 & 0 & 0 & 0 \\
			0 & 0 & 0 & 0 & 0 & 0 \\
			0 & 0 & 0 & 0 & 0 & 0 \\
			0 & 0 & 0 & 0 & 1 & 1 \\
			0 & 0 & 1 & 1 & 1 & 1 \\
			1 & 1 & 1 & 1 & 1 & 1 \\
			1 & 1 & 1 & 1 & 1 & 1 \\
			1 & 1 & 1 & 1 & 1 & 1 \\
			1 & 1 & 1 & 1 & 1 & 1 \\ \end{pmatrix}$\\
			\hline
			$E(\rho)$ && 0.45 & 0.75 & 0.9 \\
			$X$ & & $\begin{pmatrix} 0 & 0 & 0 & 0 & 0 & 0 \\
			0 & 0 & 0 & 0 & 0 & 0 \\
			0 & 0 & 0 & 0 & 0 & 0 \\
			0 & 0 & 0 & 0 & 0 & 1 \\
			0 & 0 & 0 & 0 & 1 & 1 \\
			0 & 0 & 1 & 1 & 1 & 1 \\
			0 & 1 & 1 & 1 & 1 & 1 \\
			1 & 1 & 1 & 1 & 1 & 1 \\
			1 & 1 & 1 & 1 & 1 & 1 \\
			1 & 1 & 1 & 1 & 1 & 1 \\ \end{pmatrix}$ &
			$\begin{pmatrix} 0 & 0 & 0 & 0 & 0 & 0 \\
			0 & 0 & 0 & 0 & 0 & 0 \\
			0 & 0 & 0 & 0 & 0 & 1 \\
			0 & 0 & 0 & 0 & 1 & 1 \\
			0 & 0 & 0 & 1 & 1 & 1 \\
			0 & 0 & 0 & 1 & 1 & 1 \\
			0 & 0 & 1 & 1 & 1 & 1 \\
			0 & 1 & 1 & 1 & 1 & 1 \\
			1 & 1 & 1 & 1 & 1 & 1 \\
			1 & 1 & 1 & 1 & 1 & 1 \\ \end{pmatrix}$ &
			$\begin{pmatrix} 0 & 0 & 0 & 0 & 0 & 0 \\
			0 & 0 & 0 & 0 & 0 & 1 \\
			0 & 0 & 0 & 0 & 0 & 1 \\
			0 & 0 & 0 & 0 & 1 & 1 \\
			0 & 0 & 0 & 1 & 1 & 1 \\
			0 & 0 & 0 & 1 & 1 & 1 \\
			0 & 0 & 1 & 1 & 1 & 1 \\
			0 & 1 & 1 & 1 & 1 & 1 \\
			0 & 1 & 1 & 1 & 1 & 1 \\
			1 & 1 & 1 & 1 & 1 & 1 \\ \end{pmatrix}$\\
			\bottomrule
		\end{tabular}
	\end{center}
	\caption{Optimal allocation matrices for cross-sectional designs with $D=2$. The optimal allocation matrices in the case $\mathfrak{I}=\{6\}$, $\mathfrak{C}=\{\mathfrak{C}_6\}=\{10\}$, $\mathfrak{M}=\mathfrak{M}_{6,6}=\{10\}$, and $\sigma^2=1$, with $w=0$ and $\beta=1$ are shown for a range of possible values of $E(\rho)$. No restrictions are placed on $\mathfrak{X}$ other than the identifiability of Equation~\ref{eq1}. Each allocation matrix was identified via our exhaustive search method, and matches that identified by previous research.}\label{tab1}
\end{table}

The results of our exhaustive searches are shown in Table~\ref{tab1}. In each instance the optimal design is, as would be expected, identical to that found previously. We have thus confirmed the ability of our search procedure to easily identify optimal designs for a given set of input parameters and chosen linear mixed model. Of course, in this scenario, it would likely in practice be easier to utilize the methodology of Girling and Hemming (2016).\cite{girling2016}

More recently, Thompson \textit{et al.} (2017)\cite{thompson2017} demonstrated that when $\sigma_\theta^2=\sigma_s^2=0$, if an equal number of clusters must be allocated to each sequence, then the optimal number of sequences to utilise would be
	
	\[ F(\rho)=\frac{1}{1-\sqrt{E(\rho)}}. \]
	
	We now verify their findings by restricting our set $\mathfrak{X}_{6,10}$ as follows
	
	\begin{align*}
	\mathfrak{X}_{6,10} &= \{ X : \text{dim}(X)=6\times10, X_{ij} \ge X_{ij-1} \text{ for } j=2,\dots,6 \text{ and } i=1\dots,10,\\
	& \qquad\qquad (X_{i1},\dots,X_{i6})=(X_{i'1},\dots,X_{i'6}) \text{ for } a \text{ values of } i'=1,\dots,i-1,i+1,\dots,10 \text{ and } i=1,\dots,10 \},
	\end{align*}
	
	where $a$ can be any value such that $C/a$ is an integer.
	
	For the design parameters utilised to construct Table~\ref{tab1}, we repeated our exhaustive searches but with the modified $\mathfrak{X}_{6,10}$ given above. For $E(\rho)\in\{0.10,0.15,0.30\}$ we found that the optimal $X$ was
	
	\[ X = \begin{pmatrix} 0 & 0 & 0 & 0 & 0 & 0 \\
	0 & 0 & 0 & 0 & 0 & 0 \\
	0 & 0 & 0 & 0 & 0 & 0 \\
	0 & 0 & 0 & 0 & 0 & 0 \\
	0 & 0 & 0 & 0 & 0 & 0 \\
	1 & 1 & 1 & 1 & 1 & 1 \\
	1 & 1 & 1 & 1 & 1 & 1 \\
	1 & 1 & 1 & 1 & 1 & 1 \\
	1 & 1 & 1 & 1 & 1 & 1 \\
	1 & 1 & 1 & 1 & 1 & 1 \\\end{pmatrix}. \]
	
	This should not surprise us as for $E(\rho)\in\{0.10,0.15,0.30\}$ we have $F(\rho)\in\{1.46,1.63,2.21\}$ to 2 decimal places, and the $X$ listed above is one of the few matrices belonging to the modified $\mathfrak{X}_{6,10}$ which utilises two sequences.
	
	In contrast, for $E(\rho)=0.45$, we find $F(\rho)=3.04$. However, for $C=10$ the only way equal allocation to sequences can be achieved is to utilize either two or five sequences. It should therefore not surprise us that the optimal $X$ was identified as
	
	\[ X = \begin{pmatrix}  0  &  0  &  0  &  0  &  0  &  0 \\
	0  &  0  &  0  &  0  &  0  &  0 \\
	0  &  0  &  0  &  0  &  0  &  1 \\
	0  &  0  &  0  &  0  &  0  &  1 \\
	0  &  0  &  0  &  1  &  1  &  1 \\
	0  &  0  &  0  &  1  &  1  &  1 \\
	0  &  1  &  1  &  1  &  1  &  1 \\
	0  &  1  &  1  &  1  &  1  &  1 \\
	1  &  1  &  1  &  1  &  1  &  1 \\
	1  &  1  &  1  &  1  &  1  &  1 \\\end{pmatrix}, \]
	
	which uses five sequences. Finally, for $E(\rho)\in\{0.75,0.9\}$ we have $F(\rho)\in\{7.46, 19.49\}$, and the optimal $X$ was again one which employs five sequences.

	\subsection{$D=2$: Sensitivity of the optimal designs to the variance parameter specification}\label{sens}
	
	It is important to note that our admissible design determination procedure, like the articles on optimal SW-CRTs that have come before, is dependent upon the specification of all relevant variance parameters. It is for this reason that Girling and Hemming (2016)\cite{girling2016} assessed the sensitivity of the performance of their optimised designs to the value of $E(\rho)$, via a simulation study in which $E(\rho)$ was specified using a prior.
	
\begin{figure}[htbp]
		\centerline{\includegraphics[width=0.75\linewidth]{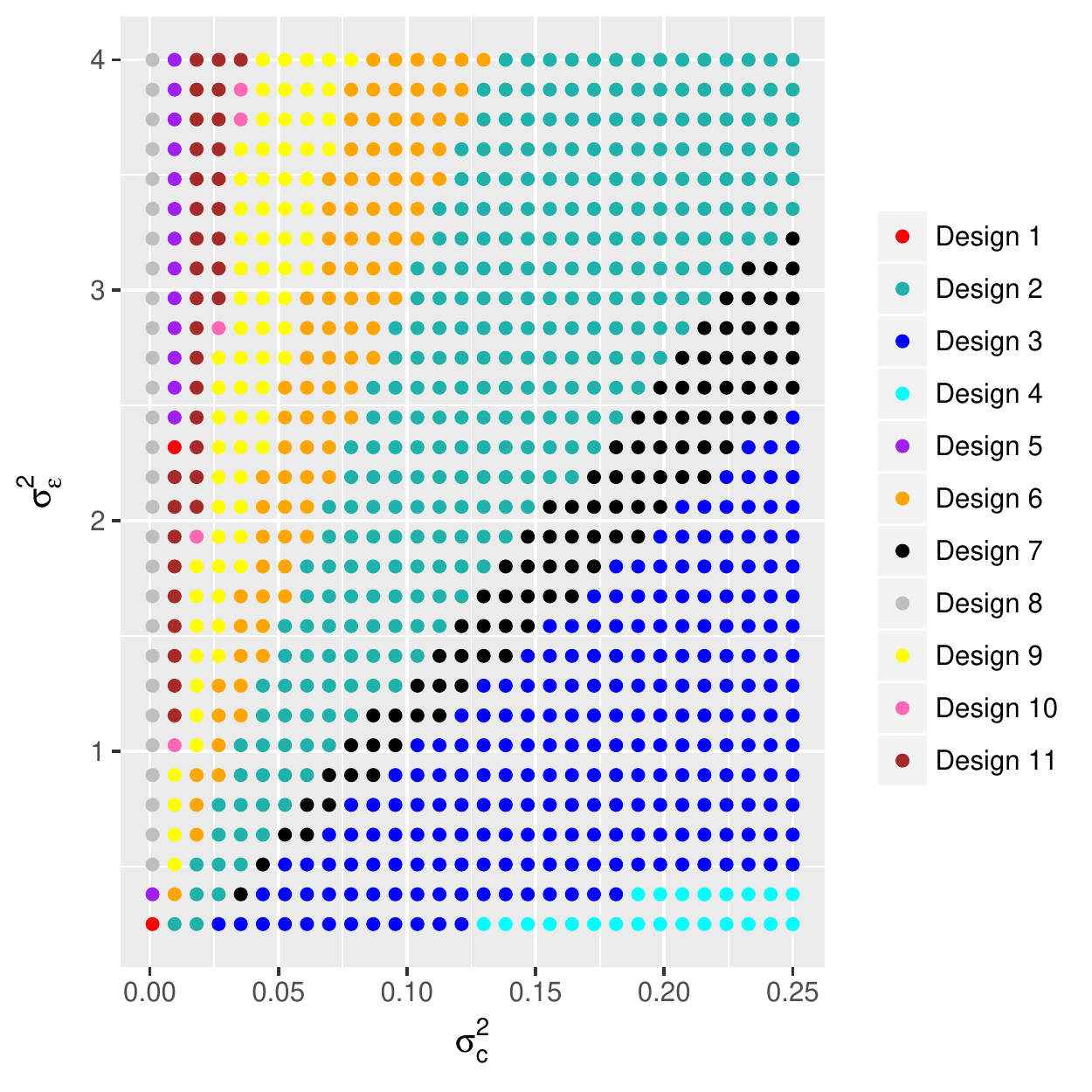}}
		\caption{Optimal allocation matrices for cross-sectional designs with $D=2$. The optimal allocation matrices in the case $\mathfrak{I}=\{6\}$, $\mathfrak{C}=\{\mathfrak{C}_6\}=\{10\}$, $\mathfrak{M}=\mathfrak{M}_{10,6}=\{10\}$, and $\sigma^2=1$, with $w=0$ and $\beta=1$ are shown for a range of possible combinations of $(\sigma_c^2,\sigma_\epsilon^2)\in[0.001,0.25]\times[0.25,4]$. No restrictions are placed on $\mathfrak{X}$ other than the identifiability of Equation~\ref{eq1}. Each allocation matrix was identified via our exhaustive search method.}\label{fig1}
	\end{figure}

	Here, we consider an alternative approach to visualising the performance of optimal designs across possible values of the variance parameters. First, in Figure~\ref{fig1}, for $w=0$, $\beta=1$, $\sigma_\theta^2=\sigma_s^2=0$, $\mathfrak{I}=\{6\}$, $\mathfrak{C}=\{\mathfrak{C}_6\}=\{10\}$, and $\mathfrak{M}=\mathfrak{M}_{10,6}=\{10\}$, we present the locations on an equally spaced grid within $(\sigma_c^2,\sigma_\epsilon^2)\in[0.001,0.25]\times[0.25,4]$ at which we identified various designs to be optimal using an exhaustive search (placing no restrictions on $\mathfrak{X}_{6,10}$). In total 11 designs were found to be optimal for at least one $(\sigma_c^2,\sigma_\epsilon^2)$ combination. We list these in full in Appendix~\ref{app1}. It would be reasonable to be troubled by this result, as it suggests a design that we believe to be optimal may not in reality be optimal if the variance parameters are even minorly misspecified.
	
\begin{figure}[t]
		\centerline{\includegraphics[width=0.7\linewidth]{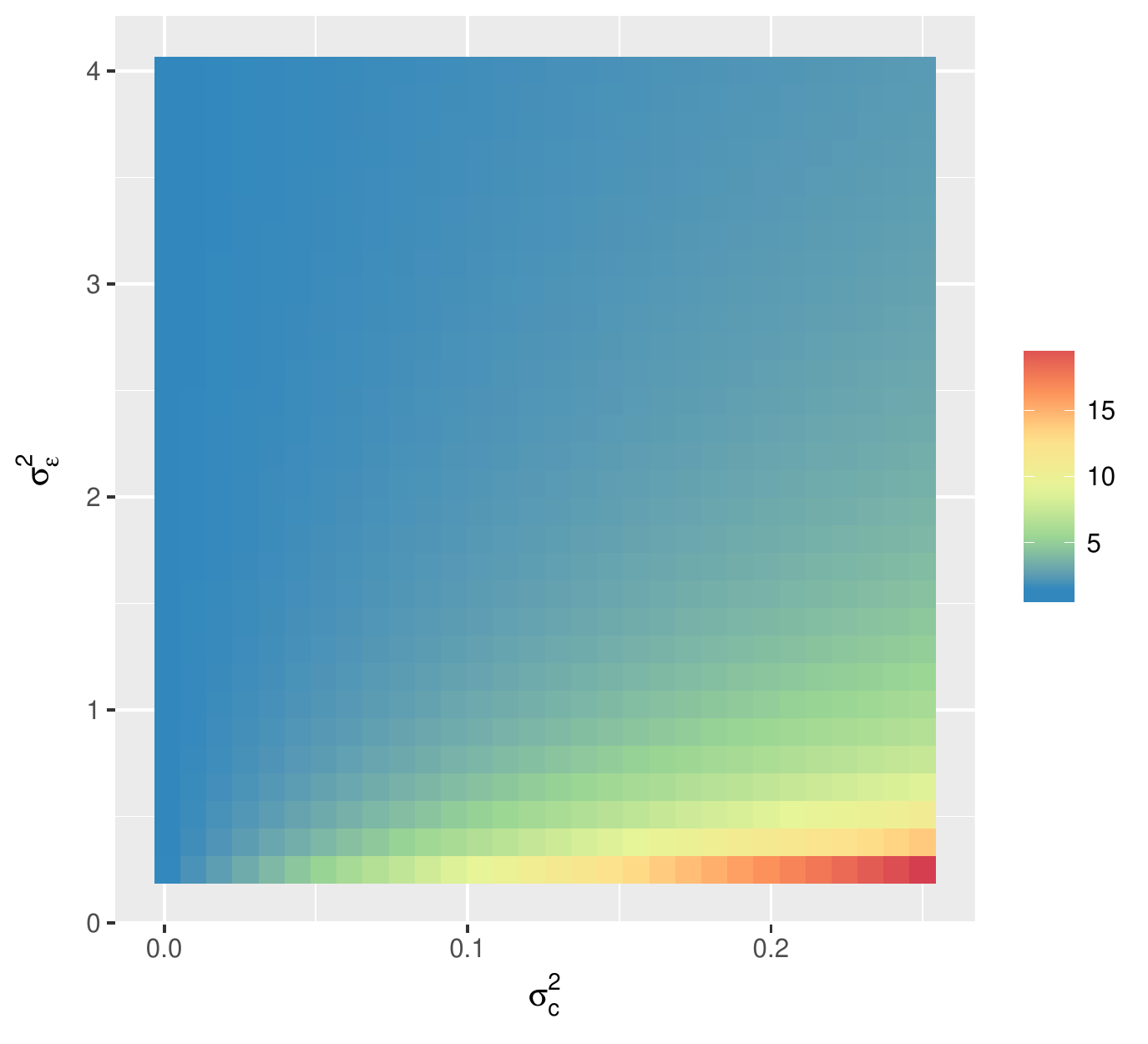}}\centerline{\includegraphics[width=0.7\linewidth]{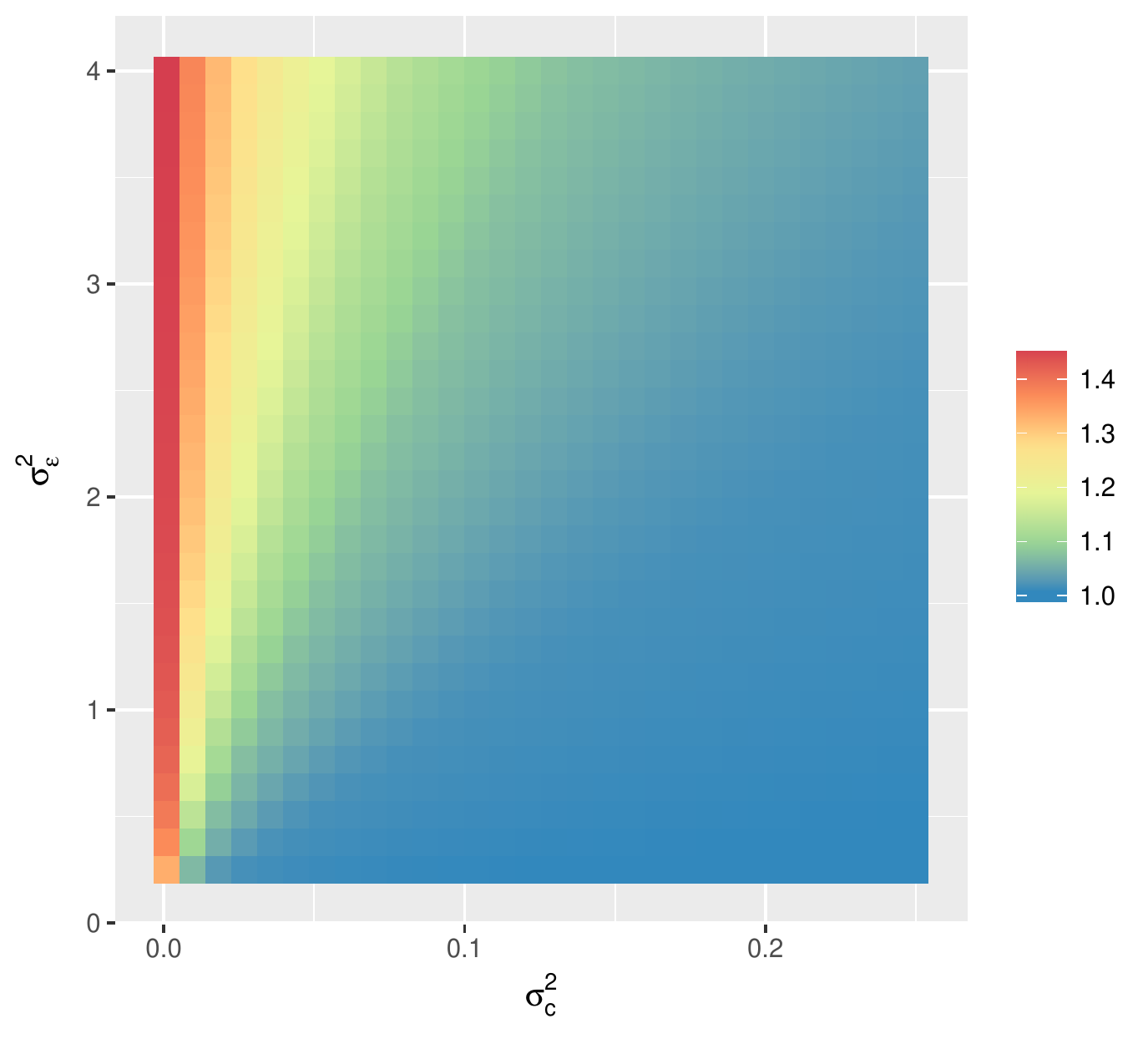}}
		\caption{The ratio of the variance of the intervention effect when using design matrices $X_1$ (top) and $X_2$ (bottom) relative to the optimal design (given in Figure~\ref{fig1}) is shown for a range of possible combinations of $(\sigma_c^2,\sigma_\epsilon^2)\in[0.001,0.25]\times[0.25,4]$.\label{fig2}}
	\end{figure}

	We can, however, inspect how large our concern should be by examining the performance of any of these optimal designs across the possible values of the variance parameters, relative to the performance of the true optimal design at each point. That is, we inspect the ratio of the variance of the intervention effect estimate of a particular design to that of the optimal design at each $(\sigma_c^2,\sigma_\epsilon^2)$ combination. We present such an evaluation in Figure~\ref{fig2} for the following design matrices
	
	\[ X_1 = \begin{pmatrix} 0 & 0 & 0 & 0 & 0 & 0 \\
	0 & 0 & 0 & 0 & 0 & 0 \\
	0 & 0 & 0 & 0 & 0 & 0 \\
	0 & 0 & 0 & 0 & 0 & 0 \\
	0 & 0 & 0 & 0 & 0 & 0 \\
	1 & 1 & 1 & 1 & 1 & 1 \\
	1 & 1 & 1 & 1 & 1 & 1 \\
	1 & 1 & 1 & 1 & 1 & 1 \\
	1 & 1 & 1 & 1 & 1 & 1 \\
	1 & 1 & 1 & 1 & 1 & 1 \\\end{pmatrix},\qquad X_2 = \begin{pmatrix} 0 & 0 & 0 & 0 & 0 & 0 \\
	0 & 0 & 0 & 0 & 0 & 1 \\
	0 & 0 & 0 & 0 & 0 & 1 \\
	0 & 0 & 0 & 0 & 1 & 1 \\
	0 & 0 & 0 & 1 & 1 & 1 \\
	0 & 0 & 0 & 1 & 1 & 1 \\
	0 & 0 & 1 & 1 & 1 & 1 \\
	0 & 1 & 1 & 1 & 1 & 1 \\
	0 & 1 & 1 & 1 & 1 & 1 \\
	1 & 1 & 1 & 1 & 1 & 1 \\ \end{pmatrix}, \]
	
	which are Designs 8 and 3 from Figure~\ref{fig1} respectively. As must obviously be the case, the value of the ratio of the variances is in all instances at least one. We observe that with the matrix $X_1$, the variance of the intervention effect estimate is substantially larger than that for the optimal design when the values of $\sigma_c^2$ and $\sigma_\epsilon^2$ are mis-specified, particularly when the value of $\rho$ is in fact large. In contrast, using the matrix $X_2$ retains efficient performance in many instances. However, if $\rho$ is small then the variance of the intervention effect provided by this design is still more than 40\% larger than that of the optimal design.
	
	\subsection{$D=2$: Li \textit{et al.} (2018)}
	
	Li \textit{et al.} (2018)\cite{li2018} recently extended the results of Lawrie \textit{et al.} (2015)\cite{lawrie2015} to cohort SW-CRTs. Specifically, they considered a case in which all clusters have to begin in the control condition (intervention 0), and conclude in the experimental (intervention 1). They then demonstrated that the optimal $X$ could be specified by ensuring that the proportion, $p_t$, of clusters allocated to a sequence with $t$ ones preceded by $T-t$ zeros satisfies
	\begin{align*}
	p_1 &= p_{T-1} = \frac{\psi+3\xi}{2\gamma},\\
	p_t &= \frac{\xi}{\gamma},\qquad \text{ for } t=2,\dots,T-2,
	\end{align*}
	where
	\begin{align*}
	\psi &= 1 - (m-1)\rho_0-(m-1)\rho_1-\rho_2,\\
	\xi &= (m-1)\rho_1+\rho_2,\\
	\gamma &= \psi + T\xi.
	\end{align*}
	
	Here, we explore their findings for several example design scenarios, again via an exhaustive search. As above, we consider the case in which $\mathfrak{I}=\{6\}$, $\mathfrak{C}=\{\mathfrak{C}_6\}=\{10\}$, and $\mathfrak{M}=\mathfrak{M}_{10,6}=\{10\}$, with $\sigma^2=1$, $w=0$, and $\beta=1$. To follow their restrictions on the allowed $X$ we enforce that
	\[ \mathfrak{X}_{6,10} = \{ X : \text{dim}(X)=6\times10, X_{ij} \ge X_{ij-1} \text{ for } j=2,\dots,6 \text{ and } i=1\dots,10, X_{i1}=0 \text{ and } X_{iT}=1\text{ for }i=1,\dots,C\}. \]
	Then, we denote by $\boldsymbol{p}_{\text{th}}=(p_1,\dots,p_{T-1})^\top$ the vector of the $p_t$ for the theoretical optimal designs derived by Li \textit{et al.} (2018),\cite{li2018} and denote by $\boldsymbol{p}_{\text{emp}}$ the vector of the empirical values of the $p_t$ for our identified optimal designs. Our findings are presented in Table~\ref{tab2} for $(\rho_0,\rho_1,\rho_2)\in\{0.05,0.1\}\times\{0.001,0.002\}\times\{0.25,0.5\}$. They illustrate one potential issue with applying the results of Li \textit{et al.} (2018)\cite{li2018} in practice; that the theoretically optimal values of the $p_t$ will likely not be achievable because $C$ is an integer. However, it is clear that the empirical values of the proportions of clusters changing to the experimental intervention in each time period are close to their theoretical values, even in this case where $C$ is small.
	
	\begin{table}[htbp]
		\renewcommand{\arraystretch}{1.2}
		\begin{center}
			\begin{small}
			\begin{tabular}{rrrrrr}
				\toprule
				Factor && \multicolumn{4}{c}{Results} \\
				\cline{1-1}\cline{3-6}
				$\rho_0$ && 0.050 & 0.050 & 0.050 & 0.050 \\
				$\rho_1$ && 0.001 & 0.001 & 0.002 & 0.002 \\
				$\rho_2$ && 0.250 & 0.500 & 0.250 & 0.500 \\
				$X$      && $\begin{pmatrix}
				0 & 0 & 0 & 0 & 0 & 1 \\
				0 & 0 & 0 & 0 & 0 & 1 \\
				0 & 0 & 0 & 0 & 0 & 1 \\
				0 & 0 & 0 & 0 & 0 & 1 \\
				0 & 0 & 0 & 0 & 1 & 1 \\
				0 & 0 & 0 & 1 & 1 & 1 \\
				0 & 0 & 1 & 1 & 1 & 1 \\
				0 & 1 & 1 & 1 & 1 & 1 \\
				0 & 1 & 1 & 1 & 1 & 1 \\
				0 & 1 & 1 & 1 & 1 & 1 \\
				\end{pmatrix}$
				& $\begin{pmatrix}
				0 & 0 & 0 & 0 & 0 & 1 \\
				0 & 0 & 0 & 0 & 0 & 1 \\
				0 & 0 & 0 & 0 & 0 & 1 \\
				0 & 0 & 0 & 0 & 1 & 1 \\
				0 & 0 & 0 & 1 & 1 & 1 \\
				0 & 0 & 0 & 1 & 1 & 1 \\
				0 & 0 & 1 & 1 & 1 & 1 \\
				0 & 1 & 1 & 1 & 1 & 1 \\
				0 & 1 & 1 & 1 & 1 & 1 \\
				0 & 1 & 1 & 1 & 1 & 1 \\
				\end{pmatrix}$
				& $\begin{pmatrix}
				0 & 0 & 0 & 0 & 0 & 1 \\
				0 & 0 & 0 & 0 & 0 & 1 \\
				0 & 0 & 0 & 0 & 0 & 1 \\
				0 & 0 & 0 & 0 & 0 & 1 \\
				0 & 0 & 0 & 0 & 1 & 1 \\
				0 & 0 & 0 & 1 & 1 & 1 \\
				0 & 0 & 1 & 1 & 1 & 1 \\
				0 & 1 & 1 & 1 & 1 & 1 \\
				0 & 1 & 1 & 1 & 1 & 1 \\
				0 & 1 & 1 & 1 & 1 & 1 \\
				\end{pmatrix}$
				& $\begin{pmatrix}
				0 & 0 & 0 & 0 & 0 & 1 \\
				0 & 0 & 0 & 0 & 0 & 1 \\
				0 & 0 & 0 & 0 & 0 & 1 \\
				0 & 0 & 0 & 0 & 1 & 1 \\
				0 & 0 & 0 & 1 & 1 & 1 \\
				0 & 0 & 0 & 1 & 1 & 1 \\
				0 & 0 & 1 & 1 & 1 & 1 \\
				0 & 1 & 1 & 1 & 1 & 1 \\
				0 & 1 & 1 & 1 & 1 & 1 \\
				0 & 1 & 1 & 1 & 1 & 1 \\
				\end{pmatrix}$\\
				$\boldsymbol{p}_{\text{th}}^\top$ && (0.30,0.08,0.08,0.08,0.30) & (0.24,0.10,0.10,0.10,0.24) & (0.29,0.08,0.08,0.08,0.29) & (0.24,0.10,0.10,0.10,0.24) \\
				$\boldsymbol{p}_{\text{emp}}^\top$ && (0.4,0.1,0.1,0.1,0.3) & (0.3,0.1,0.2,0.1,0.3) &  (0.4,0.1,0.1,0.1,0.3) & (0.3,0.1,0.2,0.1,0.3) \\
				\midrule
				$\rho_0$ && 0.100 & 0.100 & 0.100 & 0.100 \\
				$\rho_1$ && 0.001 & 0.001 & 0.002 & 0.002 \\
				$\rho_2$ && 0.250 & 0.500 & 0.250 & 0.500 \\
				$X$      && $\begin{pmatrix}
				0 & 0 & 0 & 0 & 0 & 1 \\
				0 & 0 & 0 & 0 & 0 & 1 \\
				0 & 0 & 0 & 0 & 0 & 1 \\
				0 & 0 & 0 & 0 & 0 & 1 \\
				0 & 0 & 0 & 0 & 1 & 1 \\
				0 & 0 & 1 & 1 & 1 & 1 \\
				0 & 1 & 1 & 1 & 1 & 1 \\
				0 & 1 & 1 & 1 & 1 & 1 \\
				0 & 1 & 1 & 1 & 1 & 1 \\
				0 & 1 & 1 & 1 & 1 & 1 \\
				\end{pmatrix}$
				& $\begin{pmatrix}
				0 & 0 & 0 & 0 & 0 & 1 \\
				0 & 0 & 0 & 0 & 0 & 1 \\
				0 & 0 & 0 & 0 & 0 & 1 \\
				0 & 0 & 0 & 0 & 1 & 1 \\
				0 & 0 & 0 & 1 & 1 & 1 \\
				0 & 0 & 0 & 1 & 1 & 1 \\
				0 & 0 & 1 & 1 & 1 & 1 \\
				0 & 1 & 1 & 1 & 1 & 1 \\
				0 & 1 & 1 & 1 & 1 & 1 \\
				0 & 1 & 1 & 1 & 1 & 1 \\
				\end{pmatrix}$
				& $\begin{pmatrix}
				0 & 0 & 0 & 0 & 0 & 1 \\
				0 & 0 & 0 & 0 & 0 & 1 \\
				0 & 0 & 0 & 0 & 0 & 1 \\
				0 & 0 & 0 & 0 & 0 & 1 \\
				0 & 0 & 0 & 0 & 1 & 1 \\
				0 & 0 & 1 & 1 & 1 & 1 \\
				0 & 1 & 1 & 1 & 1 & 1 \\
				0 & 1 & 1 & 1 & 1 & 1 \\
				0 & 1 & 1 & 1 & 1 & 1 \\
				0 & 1 & 1 & 1 & 1 & 1 \\
				\end{pmatrix}$
				& $\begin{pmatrix}
				0 & 0 & 0 & 0 & 0 & 1 \\
				0 & 0 & 0 & 0 & 0 & 1 \\
				0 & 0 & 0 & 0 & 0 & 1 \\
				0 & 0 & 0 & 0 & 1 & 1 \\
				0 & 0 & 0 & 1 & 1 & 1 \\
				0 & 0 & 0 & 1 & 1 & 1 \\
				0 & 0 & 1 & 1 & 1 & 1 \\
				0 & 1 & 1 & 1 & 1 & 1 \\
				0 & 1 & 1 & 1 & 1 & 1 \\
				0 & 1 & 1 & 1 & 1 & 1 \\
				\end{pmatrix}$\\
				$\boldsymbol{p}_{\text{th}}^\top$ && (0.32,0.07,0.07,0.07,0.32) & (0.26,0.10,0.10,0.10,0.26) & (0.31,0.07,0.07,0.07,0.31) & (0.26,0.10,0.10,0.10,0.26) \\
				$\boldsymbol{p}_{\text{emp}}^\top$ && (0.4,0.1,0,0.1,0.4) & (0.3,0.1,0.2,0.1,0.3) & (0.4,0.1,0,0.1,0.4) & (0.3,0.1,0.2,0.1,0.3)\\
				\bottomrule
			\end{tabular}
			\end{small}
		\end{center}
		\caption{Optimal allocation matrices for cohort designs with $D=2$. The optimal allocation matrices in the case $\mathfrak{I}=\{6\}$, $\mathfrak{C}=\{\mathfrak{C}_6\}=\{10\}$, $\mathfrak{M}=\mathfrak{M}_{10,6}=\{10\}$, and $\sigma^2=1$, with $w=0$ and $\beta=1$ are shown for a range of possible combinations of $\rho_0$, $\rho_1$, and $\rho_2$. Restrictions are placed on $\mathfrak{X}$ such that Equation~\ref{eq1} is identifiable, and that each cluster must start in the control intervention (arm 0) and conclude in the experimental intervention (arm 1). Each allocation matrix was identified via our exhaustive search method.}\label{tab2}
	\end{table}

\subsection{$D=3$: SO-HIP Study}\label{sohip}

The SO-HIP study is a cross-sectional MA-SW, with $D=3$, to evaluate the effectiveness of sensor monitoring in an occupational therapy rehabilitation program for older people after hip fracture. Specifically, arm 0 corresponds to providing participants with care as usual. Arm 1 then involves the additional use of occupational therapy without sensor monitoring, in contrast to arm 2 that incorporates occupational therapy with cognitive behavioural therapy coaching using sensor monitoring as a coaching tool. Thus, as discussed earlier, intervention $d-1$ is nested within intervention $d$, for $d=1,2$.

SO-HIP plans to enrol six clusters ($C=6$), and have six time periods ($T=6$), with eight observations made per cluster per period ($m=8$), using the following matrix for treatment allocation

\[ X = \begin{pmatrix} 0 & 0 & 0 & 1 & 1 & 2 \\
	0 & 0 & 0 & 1 & 1 & 2 \\
	0 & 0 & 1 & 1 & 2 & 2 \\
	0 & 0 & 1 & 1 & 2 & 2 \\
	0 & 1 & 1 & 2 & 2 & 2 \\
	0 & 1 & 1 & 2 & 2 & 2 \end{pmatrix}. \]
	
	The trial has $\boldsymbol{\delta}=(1.5\sigma,0.75\sigma)^\top$, and assumes that $\sigma_s^2=\sigma_\theta^2=0$ and $\rho(=\rho_0=\rho_1=\rho_2)=0.05$. With this, when $\sigma^2=1$, using our methods described above we can identify that the proposed design will have an individual power of 0.88 ($\beta=0.12$) when the familywise error-rate is controlled to $\alpha=0.05$ using the Bonferroni correction. For further information on this trial, see the published protocol.\cite{pol2017}

We now consider how much efficiency could be gained by utilizing an alternative design. We presume that in the trial any number of time periods two through six could have been employed ($\mathfrak{I}=\{2,\dots,6\}$), and any number of clusters two through six could have actually been utilized ($\mathfrak{C} = \{\mathfrak{C}_2,\dots,\mathfrak{C}_6\}$, with $\mathfrak{C}_T=\{2,\dots,6\}$ for each $T\in\mathfrak{T}$). Finally, we assume that the trials plan to recruit 48 patients in total from each cluster would allow $\mathfrak{M}_{C,T} = \{2,\dots,\lfloor 48/T \rfloor\}$. Here, we enforce that 
\[ \mathfrak{X}_{C,T} = \{ X : \text{dim}(X)=C\times T, X_{ij} \ge X_{ij-1} \text{ for } j=2,\dots,T \text{ and } i=1\dots,C\}. \]
Taking our cost function to be the total number of observations, $f(\mathscr{D}) = mCT$, we present several admissible designs in Table~\ref{tab3}. Explicitly, in this case, we find that the optimal designs when using the D-, A-, and E-optimality criteria coincide for $w=0$ and $w=0.5$. Note that we also considered the optimal designs for $w=1-10^{-4}$, but they were found to be identical to those for $w=0.5$.

We can see that the individual power of the trial could be increased by as much as 12.1\%, as a result of reducing the maximum value of the variances of the treatment effect estimators by 44.3\% ($w=0$). Alternatively, the individual power could be maintained and the required number of observations reduced by up to 58.3\% ($w=0.5$).

\begin{table}[htbp]
	\renewcommand{\arraystretch}{1.2}
	\begin{center}
		\begin{tabular}{rrrr}
			\toprule
			& \multicolumn{3}{c}{Design} \\
			\cline{2-4}
			Factor & Proposed & D/A/E-Optimal: $w=0$ & D/A/E-Optimal: $w=0.5$\\
			\midrule
			$C$ & 6 & 6 & 6 \\
			$T$ & 6 & 6 & 5 \\
			$m$ & 8 & 8 & 4 \\
			$X$ & $\begin{pmatrix} 0 & 0 & 0 & 1 & 1 & 2 \\
			0 & 0 & 0 & 1 & 1 & 2 \\
			0 & 0 & 1 & 1 & 2 & 2 \\
			0 & 0 & 1 & 1 & 2 & 2 \\
			0 & 1 & 1 & 2 & 2 & 2 \\
			0 & 1 & 1 & 2 & 2 & 2 \end{pmatrix}$ & $\begin{pmatrix} 0 & 0 & 0 & 0 & 0 & 1 \\ 0 & 0 & 0 & 0 & 1 & 1 \\ 0 & 0 & 0 & 1 & 1 & 2 \\ 0 & 1 & 1 & 2 & 2 & 2 \\ 1 & 1 & 2 & 2 & 2 & 2 \\ 1 & 2 & 2 & 2 & 2 & 2 \end{pmatrix}$ & $\begin{pmatrix} 0 & 0 & 1 & 1 & 1 \\ 0 & 0 & 1 & 1 & 1 \\ 1 & 1 & 1 & 2 & 2 \\ 1 & 1 & 2 & 2 & 2 \\ 2 & 2 & 2 & 2 & 2 \\ 2 & 2 & 2 & 2 & 2 \end{pmatrix}$ \\
			$\mathbb{P}(\text{Reject }H_{01} | \delta_1)$ & 1.000 & 1.0000 ($\pm0$\%) & 0.9937 ($-0.6$\%) \\
			$\mathbb{P}(\text{Reject }H_{02} | \delta_2)$ & 0.8815 & 0.9878 ($+12.1$\%) & 0.8818 ($\pm0$\%) \\
			$f(\mathscr{D})$ & 288 & 288 ($\pm0$\%) & 120 ($-58.3$\%) \\
			$\text{det}(\Lambda_{\mathscr{D}})$ & $3.090\times10^{-3}$ & $9.990\times10^{-4}$ ($-67.7$\%) & $6.377\times10^{-3}$ ($+106.4$\%) \\
			$(D-1)^{-1}\text{tr}(\Lambda_{\mathscr{D}})$ & $5.696\times10^{-2}$ & $3.175\times10^{-2}$ ($-44.3$\%) & $8.508\times10^{-2}$ ($+49.4$\%) \\
			$\text{max} Diag (\Lambda_{\mathscr{D}})$ & $5.696\times10^{-2}$ & $3.175\times10^{-2}$ ($-44.3$\%) & $1.132\times10^{-1}$ ($+98.8$\%) \\
			\bottomrule
		\end{tabular}
	\end{center}
	\caption{Optimal allocation matrices for cross-sectional designs with $D=3$. Several optimal allocation matrices in the case $\mathfrak{I}=\{2,\dots,6\}$, $\mathfrak{C}=\{\mathfrak{C}_2,\dots,\mathfrak{C}_6\}$, $\mathfrak{C}_T=\{2,\dots,6\}$, $\mathfrak{M}_{C,T}=\{2,\dots,\lfloor 48/T \rfloor\}$, $\sigma^2=1$, $\rho=0.05$, $\alpha=0.05$ with the Bonferroni correction, and $\beta=0.12$ for the individual power when $\boldsymbol{\delta}=(1.5\sigma,0.75\sigma)^\top$ are shown. Specifically, the optimal design for the optimality criteria is given for $w\in\{0,0.5\}$. No restrictions are placed on $\mathfrak{X}$ other than the identifiability of Equation~\ref{eq1}. Each allocation matrix was identified via our exhaustive search method. The utilized design is also shown for comparison.}\label{tab3}
\end{table}

Now, in Table~\ref{tab4}, we present corresponding evaluations, but with further restrictions placed on the sets $\mathfrak{X}_{C,T}$, as follows
	\begin{align*}
	\mathfrak{X}_{C,T} &= \{ X : \text{dim}(X)=C\times T, X_{ij} \ge X_{ij-1} \text{ for } j=2,\dots,T \text{ and } i=1\dots,C, X_{ij}=d \text{ for some } j=1,\dots,T\\&\qquad \text{ for all }d=0,\dots,D-1 \text{ and  }i=1,\dots,C\}.
	\end{align*}
	That is, we enforce that each cluster receives interventions 0, 1, and 2. This allows us to perform an assessment of the advantages optimisation can bring in the likely common case in which it is desired that each cluster receive all of the interventions. Note that in this case certain combinations of $C$ and $T$ considered above are no longer are possible (e.g., for $T=2$ a cluster cannot receive all three interventions).
	
	\begin{sidewaystable}[htbp]
		\renewcommand{\arraystretch}{1.2}
		\begin{normalsize}
			\begin{center}
				\begin{tabular}{rrrrrr}
					\toprule
					& \multicolumn{5}{c}{Design} \\
					\cline{2-6}
					Factor & Proposed & D-Optimal: $w=0$ & D-Optimal: $w=0.5$ & A/E-Optimal: $w=0$ & A/E-Optimal: $w=0.5$ \\
					\midrule
					$C$ & 6 & 6 & 6 & 6 & 6 \\
					$T$ & 6 & 6 & 6 & 6 & 6 \\
					$m$ & 8 & 8 & 5 & 8 & 5 \\
					$X$ & $\begin{pmatrix} 0 & 0 & 0 & 1 & 1 & 2 \\
					0 & 0 & 0 & 1 & 1 & 2 \\
					0 & 0 & 1 & 1 & 2 & 2 \\
					0 & 0 & 1 & 1 & 2 & 2 \\
					0 & 1 & 1 & 2 & 2 & 2 \\
					0 & 1 & 1 & 2 & 2 & 2 \end{pmatrix}$ &
					$\begin{pmatrix} 0 & 0 & 0 & 0 & 1 & 2 \\
					0 & 0 & 0 & 0 & 1 & 2 \\
					0 & 0 & 0 & 1 & 2 & 2 \\
					0 & 0 & 1 & 2 & 2 & 2 \\
					0 & 1 & 2 & 2 & 2 & 2 \\
					0 & 1 & 2 & 2 & 2 & 2 \end{pmatrix}$ & $\begin{pmatrix} 0 & 0 & 0 & 0 & 1 & 2 \\
					0 & 0 & 0 & 0 & 1 & 2 \\
					0 & 0 & 1 & 1 & 2 & 2 \\
					0 & 1 & 1 & 2 & 2 & 2 \\
					0 & 1 & 2 & 2 & 2 & 2 \\
					0 & 1 & 2 & 2 & 2 & 2 \end{pmatrix}$ & $\begin{pmatrix} 0 & 0 & 0 & 0 & 1 & 2 \\
					0 & 0 & 0 & 0 & 1 & 2 \\
					0 & 0 & 1 & 1 & 2 & 2 \\
					0 & 0 & 1 & 2 & 2 & 2 \\
					0 & 1 & 2 & 2 & 2 & 2 \\
					0 & 1 & 2 & 2 & 2 & 2 \end{pmatrix}$ & $\begin{pmatrix} 0 & 0 & 0 & 0 & 1 & 2 \\
					0 & 0 & 0 & 0 & 1 & 2 \\
					0 & 0 & 1 & 1 & 2 & 2 \\
					0 & 0 & 1 & 2 & 2 & 2 \\
					0 & 1 & 2 & 2 & 2 & 2 \\
					0 & 1 & 2 & 2 & 2 & 2 \end{pmatrix}$ \\
					$\mathbb{P}(\text{Reject }H_{01} | \delta_1)$ & 1.0000 & 1.0000 ($\pm0$\%) & 1.0000 ($\pm0$\%) & 1.0000 ($\pm0$\%) & 1.0000 ($\pm0$\%) \\
					$\mathbb{P}(\text{Reject }H_{02} | \delta_2)$ & 0.8815 & 0.9528 ($+8.1$\%) & 0.8507 ($-3.5$\%) & 0.9570 ($+8.6$\%) & 0.8440 ($-4.3$\%) \\
					$f(\mathscr{D})$ & 288 & 288 ($\pm0$\%) & 180 ($-37.5$\%) & 288 ($\pm0$\%) & 180 ($-37.5$\%) \\
					$\text{det}(\Lambda_{\mathscr{D}})$ & $3.090\times10^{-3}$ & $1.670\times10^{-3}$ ($-46.0$\%) &  $3.881\times10^{-3}$ ($+25.6$\%) & $1.712\times10^{-3}$ ($-44.6$\%) & $3.973\times10^{-3}$ ($+25.6$\%) \\
					$(D-1)^{-1}\text{tr}(\Lambda_{\mathscr{D}})$ & $5.696\times10^{-2}$ & $4.264\times10^{-2}$ ($-25.1$\%) & $6.392\times10^{-2}$ ($+12.2$\%) & $4.160\times10^{-2}$ ($-27.0$\%) & $6.373\times10^{-2}$ ($+11.9$\%) \\
					$\text{max} Diag (\Lambda_{\mathscr{D}})$ & $5.696\times10^{-2}$ & $4.264\times10^{-2}$ ($-25.1$\%) & $6.531\times10^{-2}$ ($+14.7$\%) & $4.160\times10^{-2}$ ($-27.0$\%) & $6.373\times10^{-2}$ ($+11.9$\%) \\
					\bottomrule
				\end{tabular}
			\end{center}
		\end{normalsize}
		\caption{Optimal allocation matrices for cross-sectional designs with $D=3$. Several optimal allocation matrices in the case $\mathfrak{I}=\{2,\dots,6\}$, $\mathfrak{C}=\{\mathfrak{C}_2,\dots,\mathfrak{C}_6\}$, $\mathfrak{C}_T=\{2,\dots,6\}$, $\mathfrak{M}_{C,T}=\{2,\dots,\lfloor 48/T \rfloor\}$, $\sigma^2=1$, $\rho=0.05$, $\alpha=0.05$ with the Bonferroni correction, and $\beta=0.12$ for the individual power when $\boldsymbol{\delta}=(1.5\sigma,0.75\sigma)^\top$ are shown. Specifically, the optimal design for the optimality criteria is given for $w\in\{0,0.5\}$. Restrictions are placed on $\mathfrak{X}$ such that Equation~\ref{eq1} is identifiable, and that each cluster must receive each of the interventions. Each allocation matrix was identified via our exhaustive search method. The utilized design is also shown for comparison.}\label{tab4}
	\end{sidewaystable}
	
	We now find that whilst the optimal designs are equivalent when using the A- or E-optimality criteria, the D-optimal designs are distinct. Overall, while the potential efficiency gains that are possible when restricting to these more classical designs are more modest than those in Table~\ref{tab3}, they are still substantial. In particular, the admissible designs with $w=0.5$ provide a 37.5\% reduction in the required number of observations compared to the utilised design. Moreover, we can still increase the individual power by up to 8.6\%.

	\subsection{$D=3$: Optimal cross-sectional designs according to the value of the cluster mean correlation}\label{across}
	
	We have now noted the fact that previous papers have described how the optimal cross-sectional SW-CRT design when $D=2$ changes according to the value of the cluster mean correlation $E(\rho)$ (where $\rho=\rho_0=\rho_1=\rho_2$ for $\sigma_s^2=\sigma_\theta^2=0$). In fact, in Table~\ref{tab1} we provide an example of this for a case with $C=10$ and $T=6$. In it, we observe that the optimal design as $E(\rho)$ increases changes from one resembling a parallel group CRT, to a more classical SW-CRT design. Here, we provide a brief assessment of whether such a pattern exists for designs with $D=3$, in a setting motivated by the SO-HIP trial. Thus, we set $\mathfrak{I}=\{6\}$, $\mathfrak{C}=\{\mathfrak{C}_6\}=\{6\}$, $\mathfrak{M}=\mathfrak{M}_{6,6}=\{8\}$, $\sigma^2=1$, $w=0$, $\beta=1$ and
	
	\[ \mathfrak{X}_{6,6} = \{ X : \text{dim}(X)=6\times 6, X_{ij} \ge X_{ij-1} \text{ for } j=2,\dots,6 \text{ and } i=1\dots,6\}. \]
	
	We then consider which design is optimal according to the D-, A-, and E-optimality criteria for $E(\rho)\in\{0,0.01,\dots,1\}$. We present our findings for E-optimality in Table~\ref{tab5}, and for D- and A-optimality in Appendix~\ref{appopt}. Specifically we can see that whilst the pattern to the way in which the optimal $X$ changes is arguably less clear than in the case with $D=2$, there is still a trend that the best possible choice shifts from a longitudinal parallel group CRT, to a design resembling an extension of a classical SW-CRT.
	
	\begin{sidewaystable}[htbp]
		\renewcommand{\arraystretch}{1.2}
		\begin{center}
			\begin{tabular}{rccccccc}
				\toprule
				Factor && \multicolumn{6}{c}{E-optimal designs} \\
				\cline{1-1}\cline{3-8}
				$E(\rho)$ && $\{0,\dots,0.06\}$ & 0.07 & $\{0.08,\dots,0.11\}$ & $\{0.12,\dots,0.34\}$ & $\{0.35,0.36\}$ & $\{0.37,\dots,0.65\}$ \\
				$X$ && $\begin{pmatrix} 0 & 0 & 0 & 0 & 0 & 0 \\
				0 & 0 & 0 & 0 & 0 & 1 \\
				1 & 1 & 1 & 1 & 1 & 1 \\
				1 & 1 & 1 & 1 & 1 & 1 \\
				1 & 2 & 2 & 2 & 2 & 2 \\
				2 & 2 & 2 & 2 & 2 & 2 \end{pmatrix}$ & $\begin{pmatrix} 0 & 0 & 0 & 0 & 0 & 0 \\
				0 & 0 & 0 & 0 & 1 & 1 \\
				1 & 1 & 1 & 1 & 1 & 1 \\
				1 & 1 & 1 & 1 & 1 & 1 \\
				1 & 1 & 2 & 2 & 2 & 2 \\
				2 & 2 & 2 & 2 & 2 & 2 \end{pmatrix}$ & $\begin{pmatrix} 0 & 0 & 0 & 0 & 0 & 0 \\
				0 & 0 & 0 & 0 & 1 & 1 \\
				0 & 1 & 1 & 1 & 1 & 1 \\
				1 & 1 & 1 & 1 & 1 & 2 \\
				1 & 1 & 2 & 2 & 2 & 2 \\
				2 & 2 & 2 & 2 & 2 & 2 \end{pmatrix}$ & $\begin{pmatrix} 0 & 0 & 0 & 0 & 0 & 0 \\
				0 & 0 & 0 & 0 & 1 & 1 \\
				0 & 0 & 1 & 1 & 1 & 1 \\
				1 & 1 & 1 & 1 & 2 & 2 \\
				1 & 1 & 2 & 2 & 2 & 2 \\
				2 & 2 & 2 & 2 & 2 & 2 \end{pmatrix}$ & $\begin{pmatrix} 0 & 0 & 0 & 0 & 0 & 1 \\
				0 & 0 & 0 & 0 & 1 & 1 \\
				0 & 0 & 1 & 1 & 1 & 1 \\
				1 & 1 & 1 & 1 & 2 & 2 \\
				1 & 1 & 2 & 2 & 2 & 2 \\
				1 & 2 & 2 & 2 & 2 & 2 \end{pmatrix}$ & $\begin{pmatrix} 0 & 0 & 0 & 0 & 0 & 1 \\
				0 & 0 & 0 & 0 & 1 & 1 \\
				0 & 0 & 1 & 1 & 1 & 2 \\
				0 & 1 & 1 & 1 & 2 & 2 \\
				1 & 1 & 2 & 2 & 2 & 2 \\
				1 & 2 & 2 & 2 & 2 & 2 \end{pmatrix}$\\
				\midrule
				$E(\rho)$ && $\{0.66,\dots,0.83\}$ & 0.84 & 0.85 & $\{0.86,\dots,0.94\}$ & $\{0.95,\dots,0.99\}$ & 1.00 \\
				$X$ && $\begin{pmatrix} 0 & 0 & 0 & 0 & 0 & 1 \\
				0 & 0 & 0 & 0 & 1 & 1 \\
				0 & 0 & 0 & 1 & 1 & 2 \\
				0 & 1 & 1 & 2 & 2 & 2 \\
				1 & 1 & 2 & 2 & 2 & 2 \\
				1 & 2 & 2 & 2 & 2 & 2 \end{pmatrix}$ & $\begin{pmatrix} 0 & 0 & 0 & 0 & 0 & 1 \\
				0 & 0 & 0 & 0 & 1 & 1 \\
				0 & 0 & 0 & 1 & 1 & 2 \\
				0 & 1 & 2 & 2 & 2 & 2 \\
				1 & 1 & 1 & 2 & 2 & 2 \\
				1 & 2 & 2 & 2 & 2 & 2 \end{pmatrix}$ & $\begin{pmatrix} 0 & 0 & 0 & 0 & 0 & 1 \\
				0 & 0 & 0 & 0 & 1 & 2 \\
				0 & 0 & 0 & 1 & 1 & 1 \\
				0 & 1 & 1 & 2 & 2 & 2 \\
				1 & 1 & 2 & 2 & 2 & 2 \\
				1 & 2 & 2 & 2 & 2 & 2 \end{pmatrix}$ & $\begin{pmatrix} 0 & 0 & 0 & 0 & 0 & 1 \\
				0 & 0 & 0 & 0 & 1 & 1 \\
				0 & 0 & 0 & 1 & 2 & 2 \\
				0 & 0 & 1 & 2 & 2 & 2 \\
				1 & 1 & 2 & 2 & 2 & 2 \\
				1 & 2 & 2 & 2 & 2 & 2 \end{pmatrix}$ & $\begin{pmatrix} 0 & 0 & 0 & 0 & 0 & 1 \\
				0 & 0 & 0 & 0 & 1 & 2 \\
				0 & 0 & 0 & 1 & 2 & 2 \\
				0 & 0 & 1 & 2 & 2 & 2 \\
				0 & 1 & 2 & 2 & 2 & 2 \\
				1 & 2 & 2 & 2 & 2 & 2 \end{pmatrix}$ & $\begin{pmatrix} 0 & 0 & 0 & 0 & 0 & 0 \\
				0 & 0 & 0 & 0 & 0 & 0 \\
				0 & 0 & 0 & 0 & 0 & 0 \\
				1 & 2 & 2 & 2 & 2 & 2 \\
				1 & 2 & 2 & 2 & 2 & 2 \\
				1 & 2 & 2 & 2 & 2 & 2 \end{pmatrix}$ \\
				\bottomrule
			\end{tabular}
		\end{center}
		\caption{E-optimal allocation matrices for cross-sectional designs with $D=3$. The E-optimal allocation matrices in the case $\mathfrak{I}=\{6\}$, $\mathfrak{C}=\{\mathfrak{C}_6\}=\{6\}$, $\mathfrak{M}=\mathfrak{M}_{6,6}=\{8\}$, and $\sigma^2=1$, with $w=0$ and $\beta=1$ are shown for $E(\rho)\in\{0,0.01,\dots,1\}$. No restrictions are placed on $\mathfrak{X}$ other than the identifiability of Equation~\ref{eq1}. Each allocation matrix was identified via our exhaustive search method.}\label{tab5}
	\end{sidewaystable}
	
	\subsection{$D=4$: Stochastic determination of optimal designs}
	
	Finally, we suppose that the SO-HIP study is to actually be conducted with a fourth intervention arm. This hypothetical trial is to again be conducted in six clusters ($C=6$), with eight measurements taken per cluster per period ($m=8$), but will now run across eight periods ($T=8$). Furthermore, the following natural extension of the design for $D=3$ will be used for $X$
	
	\[ X = \begin{pmatrix} 0 & 0 & 0 & 1 & 1 & 2 & 2 & 3 \\
	0 & 0 & 0 & 1 & 1 & 2 & 2 & 3 \\
	0 & 0 & 1 & 1 & 2 & 2 & 3 & 3 \\
	0 & 0 & 1 & 1 & 2 & 2 & 3 & 3 \\
	0 & 1 & 1 & 2 & 2 & 3 & 3 & 3 \\
	0 & 1 & 1 & 2 & 2 & 3 & 3 & 3 \end{pmatrix}. \]

We assume that the trial will control the familywise error-rate to $\alpha=0.05$ using the Bonferroni correction. Pre-trial, the variance parameters have been set as $\sigma^2=1$ and $\rho=0.05$, and we take $\boldsymbol{\delta}=(1.5\sigma,0.75\sigma,0.75\sigma)^\top$.

We then suppose that we desire to determine how much the trials efficiency could be improved if an alternative design was utilized. For this we employ a stochastic search, as $\mathfrak{I}=\{8\}$, $\mathfrak{C}=\{\mathfrak{C}_8\}=\{6\}$, and $\mathfrak{M}=\{ \mathfrak{M}_{6,8} \} = \{8\}$ with $D=4$ confer a design space too large for an exhaustive comparison.

In Table~\ref{tab6} we present the stochastically identified optimal designs for the D-, A-, and E-optimality crtieria. We can see that, in particular, the average variance of our intervention effects could be reduced by up to 49.8\% (A-optimality), or the maximal variance of the intervention effects reduced by up to 48.2\% (E-optimality). It is thus clear that a stochastic search can allow the identification of efficient designs when an exhaustive search would not be feasible. 

\begin{sidewaystable}[htbp]
	\renewcommand{\arraystretch}{1.2}
	\begin{center}
		\begin{tabular}{rrrrr}
			\toprule
			& \multicolumn{4}{c}{Design} \\
			\cline{2-5}
			Factor & Proposed & D-optimal & A-optimal & E-optimal \\
			\midrule
			$X$ & $\begin{pmatrix} 0 & 0 & 0 & 1 & 1 & 2 & 2 & 3 \\
			0 & 0 & 0 & 1 & 1 & 2 & 2 & 3 \\
			0 & 0 & 1 & 1 & 2 & 2 & 3 & 3 \\
			0 & 0 & 1 & 1 & 2 & 2 & 3 & 3 \\
			0 & 1 & 1 & 2 & 2 & 3 & 3 & 3 \\
			0 & 1 & 1 & 2 & 2 & 3 & 3 & 3 \end{pmatrix}$ & $\begin{pmatrix} 0 & 0 & 0 & 0 & 0 & 0 & 1 & 1 \\
			0 & 0 & 0 & 0 & 1 & 1 & 2 & 3 \\
			0 & 0 & 1 & 2 & 2 & 3 & 3 & 3 \\
			0 & 1 & 1 & 1 & 1 & 2 & 2 & 2 \\
			1 & 2 & 2 & 2 & 3 & 3 & 3 & 3 \\
			2 & 2 & 3 & 3 & 3 & 3 & 3 & 3 \end{pmatrix}$ & $\begin{pmatrix} 0 & 0 & 0 & 0 & 0 & 0 & 1 & 1 \\
			0 & 0 & 0 & 0 & 1 & 1 & 3 & 3 \\
			0 & 0 & 1 & 1 & 1 & 2 & 2 & 2 \\
			1 & 1 & 1 & 1 & 2 & 2 & 2 & 2 \\
			1 & 1 & 2 & 2 & 2 & 3 & 3 & 3 \\
			2 & 2 & 2 & 3 & 3 & 3 & 3 & 3 \end{pmatrix}$ & $\begin{pmatrix} 0 & 0 & 0 & 0 & 0 & 0 & 1 & 3 \\
			0 & 0 & 0 & 0 & 1 & 1 & 2 & 2 \\
			0 & 0 & 0 & 1 & 1 & 3 & 3 & 3 \\
			1 & 1 & 1 & 1 & 2 & 2 & 2 & 2 \\
			1 & 1 & 2 & 2 & 3 & 3 & 3 & 3 \\
			2 & 2 & 3 & 3 & 3 & 3 & 3 & 3 \end{pmatrix}$ \\
			$\mathbb{P}(\text{Reject }H_{01} | \delta_1)$ & 1.000 & 1.000 ($\pm0$\%) & 1.000 ($\pm0$\%) & 1.000 ($\pm0$\%)\\
			$\mathbb{P}(\text{Reject }H_{02} | \delta_2)$ & 0.852 & 0.992 ($+11.6$\%) & 0.996 ($+11.7$\%) & 0.989 ($+11.6$\%) \\
			$\mathbb{P}(\text{Reject }H_{03} | \delta_3)$ & 0.852 & 0.990 ($+11.6$\%) & 0.984 ($+11.6$\%) & 0.989 ($+11.6$\%) \\
			$\text{det}(\Lambda_{\mathscr{D}})$ & $1.559\times10^{-4}$ & $1.985\times10^{-5}$ ($-87.3$\%) & $2.108\times10^{-5}$ ($-86.5$\%) & $2.090\times10^{-5}$ ($-86.6$\%) \\
			$(D-1)^{-1}\text{tr}(\Lambda_{\mathscr{D}})$ & $5.590\times10^{-2}$ & $2.873\times10^{-2}$ ($-48.6$\%) & $2.806\times10^{-2}$ ($-49.8$\%) & $2.886\times10^{-2}$ ($-48.4$\%) \\
			$\text{max} Diag (\Lambda_{\mathscr{D}})$ & $5.590\times10^{-2}$ & $3.024\times10^{-2}$ ($-45.9$\%) & $3.085\times10^{-2}$ ($-44.8$\%) & $2.893\times10^{-2}$ ($-48.2$\%) \\
			\bottomrule
		\end{tabular}
	\end{center}
	\caption{Optimal allocation matrices for cross-sectional designs with $D=4$. Several optimal allocation matrices in the case $\mathfrak{I}=\{8\}$, $\mathfrak{C}=\{\mathfrak{C}_6\}=\{8\}$, $\mathfrak{M}_{6,8}=\{8\}$, $\sigma^2=1$, $\rho=0.05$, $\alpha=0.05$ with the Bonferroni correction, $w=0$, and $\beta=0.12$ for the individual power when $\boldsymbol{\delta}=(1.5\sigma,0.75\sigma,0.75\sigma)^\top$, are shown. No restrictions are placed on $\mathfrak{X}$ other than the identifiability of Equation~\ref{eq1}. Each allocation matrix was identified via our stochastic search method. The proposed design is also shown for comparison.}\label{tab6}
\end{sidewaystable}

\section{Discussion}\label{disc}

We have presented a method to determine admissible MA-SW designs. Our work builds on previous results for SW-CRTs to allow trialists to determine efficient designs when any linear mixed model is to be used for data analysis, and when there is any number of treatment arms.

For our primary motivating example, the SO-HIP study, we demonstrated for the considered parameters that the individual power could have been maintained with the number of required observations reduced by 58\%. Whilst for some possible design parameter combinations this reduction would likely not be so pronounced, it is clear that admissible designs in this context could bring notable efficiency gains.

It is important to note, however, that there are some scenarios in which our approach would likely not be applicable. This includes cases where the design space $\mathfrak{D}$ is extremely large, even after $C$ and $T$ have been specified precisely. A trialist must then either look to extend the approach of Girling and Hemming (2016),\cite{girling2016} or look to reduce the size of $\mathfrak{D}$ to make an exhaustive or stochastic search possible.

More significantly, our methodology, like all others on optimal SW-CRT design, assumes that the variance parameters of the analysis model of interest are known. Accordingly, our approach may not be a wise one when substantial uncertainty exists about their values. When confidence does exist around their specification, it remains important to assess the sensitivity of the chosen design to the underlying assumptions, using for example an approach like that in Section~\ref{sens}.

Our methodology is also limited to linear mixed models, and assumes that the employed analysis model is appropriate for the trial's data. For large sample sizes our methods may still be appropriate for alternate endpoints such as binary or count data, but they would not always be acceptable in these domains. In Appendix~\ref{app2} we provide a brief demonstration of how our methods can be applied to binary outcome variables. In addition, for some linear mixed models, allowing the number of time periods $T$ to vary may cause issues if a complex correlation structure is assumed for the accrued responses. As for any trial, the analysis model should be chosen carefully, as the chosen design may not be optimal for an alternative potential model. However, we highlight again that our approach is applicable to any linear mixed model. Thus, more complex models than that considered here are supported, including for example those which allow for the decay of treatment effects over time.

We made few principal assumptions about the nature of the trial design. Our method is applicable to both cross-sectional and cohort studies, and to cases where either a single or multiple interventions are allocated to each cluster in each time period. Nonetheless, from those MA-SW trials conducted so far, it appears that a common likelihood will be that there is some natural ordering to the interventions. Lyons \textit{et al.} (2017),\cite{lyons2017} however, do provide a detailed description of alternative possibilities to this.
	
	In Section~\ref{results} we employed several different types of restrictions on the sets $\mathfrak{X}_{C,T}$. In particular, we demonstrated our approach can be easily applied to attain classical designs where the clusters receive all interventions, and to cases where there must be equal allocation to sequences. In general, not placing restrictions on $\mathfrak{X}_{C,T}$, beyond those which are absolutely required, will result in the determination of the most efficient design. However, particularly through Table~\ref{tab4}, we were able to demonstrate that optimisation is still useful when such restrictions are considered necessary.
	
	Finally, it is important to discuss the fact that in practice a choice must be made around which optimality criteria to use, and what value to use for $w$. Unfortunately, there is no simple solution to this. Previous authors have highlighted that D-optimality is an easy quantity to explain to practitioners from many fields.\cite{dmitrienko2007} However, it is difficult to claim that A- and E-optimality would be more complex to describe. Arguably, A-optimality is most useful when the parameters of interest are of equal importance. In contrast, D- and E-optimality may favour more specialised considerations. However, note that in certain situations, as in Table~\ref{tab1}, we may find that the optimal design for each of these criteria is equivalent. Thus, such a choice may not always be required. Finally, when choosing $w$, if gathering observations is cheap we may anticipate that setting $w$ approximately equal to 0 is logical. This would also be the case when we have a fixed number of observations in mind, and simply want to optimize $X$, as in many of the discussions in Section~\ref{results}. Most typically though, it is likely we would need to find a balance between cost and efficiency. In this case, larger values of $w$ would seem appealing. But, we would rarely recommend setting $w=1$, as even placing a tiny weight on the D-, A-, or E- optimality criteria can result in the choice of a much more efficient $X$, for only slightly increased cost.

In conclusion, we have presented methodology to identify highly efficient MA-SWs. Of course, the most important factor for any real trial is that a design and analysis procedure are chosen that are appropriate for the complexities of the data the trial will likely accrue. However, when logistical, practical, and statistical, constraints permit the possibility to use one of a range of designs, researchers should consider the use of our approach to optimize their trials efficiency. As we have demonstrated, restrictions can readily be placed on the sets $\mathfrak{X}_{C,T}$ to retain the needs of the trial, but still allow more efficient designs to be identified.

\section*{Acknowledgements}

This work was supported by the Medical Research Council [grant number MC\_UP\_1302/2 to APM and MJG]; and the National Institute for Health Research Cambridge Biomedical Research Centre [MC\_UP\_1302/6 to JMSW].

\appendix

	\section{Optimal cross-sectional designs from Section 3.2}\label{app1}
	
	In Table~\ref{tab7} we list the optimal designs from Figure~\ref{fig1}, discussed in Section~\ref{sens}.
	
	\begin{table}[htbp]
		\renewcommand{\arraystretch}{1.2}
		\begin{center}
			\begin{tabular}{cccc}
				\toprule
				Design 1 & Design 2 & Design 3 & Design 4 \\
				\midrule
				$\begin{pmatrix} 0 & 0 & 0 & 0 & 0 & 0 \\
				0 & 0 & 0 & 0 & 0 & 0 \\
				0 & 0 & 0 & 0 & 0 & 0 \\
				0 & 0 & 0 & 0 & 0 & 0 \\
				0 & 0 & 0 & 0 & 1 & 1 \\
				0 & 1 & 1 & 1 & 1 & 1 \\
				1 & 1 & 1 & 1 & 1 & 1 \\
				1 & 1 & 1 & 1 & 1 & 1 \\
				1 & 1 & 1 & 1 & 1 & 1 \\
				1 & 1 & 1 & 1 & 1 & 1
				\end{pmatrix}$ &
				$\begin{pmatrix} 0 & 0 & 0 & 0 & 0 & 0 \\
				0 & 0 & 0 & 0 & 0 & 0 \\
				0 & 0 & 0 & 0 & 0 & 1 \\
				0 & 0 & 0 & 0 & 1 & 1 \\
				0 & 0 & 0 & 1 & 1 & 1 \\
				0 & 0 & 0 & 1 & 1 & 1 \\
				0 & 0 & 1 & 1 & 1 & 1 \\
				0 & 1 & 1 & 1 & 1 & 1 \\
				1 & 1 & 1 & 1 & 1 & 1 \\
				1 & 1 & 1 & 1 & 1 & 1
				\end{pmatrix}$ & 
				$\begin{pmatrix} 0 & 0 & 0 & 0 & 0 & 0 \\
				0 & 0 & 0 & 0 & 0 & 1 \\
				0 & 0 & 0 & 0 & 0 & 1 \\
				0 & 0 & 0 & 0 & 1 & 1 \\
				0 & 0 & 0 & 1 & 1 & 1 \\
				0 & 0 & 0 & 1 & 1 & 1 \\
				0 & 0 & 1 & 1 & 1 & 1 \\
				0 & 1 & 1 & 1 & 1 & 1 \\
				0 & 1 & 1 & 1 & 1 & 1 \\
				1 & 1 & 1 & 1 & 1 & 1
				\end{pmatrix}$ &
				$\begin{pmatrix} 0 & 0 & 0 & 0 & 0 & 0 \\
				0 & 0 & 0 & 0 & 0 & 1 \\
				0 & 0 & 0 & 0 & 1 & 1 \\
				0 & 0 & 0 & 0 & 1 & 1 \\
				0 & 0 & 0 & 1 & 1 & 1 \\
				0 & 0 & 0 & 1 & 1 & 1 \\
				0 & 0 & 1 & 1 & 1 & 1 \\
				0 & 1 & 1 & 1 & 1 & 1 \\
				0 & 1 & 1 & 1 & 1 & 1 \\
				1 & 1 & 1 & 1 & 1 & 1
				\end{pmatrix}$\\
				\midrule
				Design 5 & Design 6 & Design 7 & Design 8 \\
				\midrule
				$\begin{pmatrix} 0 & 0 & 0 & 0 & 0 & 0 \\
				0 & 0 & 0 & 0 & 0 & 0 \\
				0 & 0 & 0 & 0 & 0 & 0 \\
				0 & 0 & 0 & 0 & 0 & 0 \\
				0 & 0 & 0 & 0 & 0 & 1 \\
				0 & 1 & 1 & 1 & 1 & 1 \\
				1 & 1 & 1 & 1 & 1 & 1 \\
				1 & 1 & 1 & 1 & 1 & 1 \\
				1 & 1 & 1 & 1 & 1 & 1 \\
				1 & 1 & 1 & 1 & 1 & 1
				\end{pmatrix}$ &
				$\begin{pmatrix} 0 & 0 & 0 & 0 & 0 & 0 \\
				0 & 0 & 0 & 0 & 0 & 0 \\
				0 & 0 & 0 & 0 & 0 & 1 \\
				0 & 0 & 0 & 0 & 1 & 1 \\
				0 & 0 & 0 & 1 & 1 & 1 \\
				0 & 0 & 1 & 1 & 1 & 1 \\
				0 & 1 & 1 & 1 & 1 & 1 \\
				1 & 1 & 1 & 1 & 1 & 1 \\
				1 & 1 & 1 & 1 & 1 & 1 \\
				1 & 1 & 1 & 1 & 1 & 1
				\end{pmatrix}$ & 
				$\begin{pmatrix} 0 & 0 & 0 & 0 & 0 & 0 \\
				0 & 0 & 0 & 0 & 0 & 1 \\
				0 & 0 & 0 & 0 & 0 & 1 \\
				0 & 0 & 0 & 0 & 1 & 1 \\
				0 & 0 & 0 & 1 & 1 & 1 \\
				0 & 0 & 1 & 1 & 1 & 1 \\
				0 & 0 & 1 & 1 & 1 & 1 \\
				0 & 1 & 1 & 1 & 1 & 1 \\
				1 & 1 & 1 & 1 & 1 & 1 \\
				1 & 1 & 1 & 1 & 1 & 1
				\end{pmatrix}$ &
				$\begin{pmatrix} 0 & 0 & 0 & 0 & 0 & 0 \\
				0 & 0 & 0 & 0 & 0 & 0 \\
				0 & 0 & 0 & 0 & 0 & 0 \\
				0 & 0 & 0 & 0 & 0 & 0 \\
				0 & 0 & 0 & 0 & 0 & 0 \\
				1 & 1 & 1 & 1 & 1 & 1 \\
				1 & 1 & 1 & 1 & 1 & 1 \\
				1 & 1 & 1 & 1 & 1 & 1 \\
				1 & 1 & 1 & 1 & 1 & 1 \\
				1 & 1 & 1 & 1 & 1 & 1
				\end{pmatrix}$\\
				\midrule
				Design 9 & Design 10 & Design 11 & \\
				\midrule
				$\begin{pmatrix} 0 & 0 & 0 & 0 & 0 & 0 \\
				0 & 0 & 0 & 0 & 0 & 0 \\
				0 & 0 & 0 & 0 & 0 & 0 \\
				0 & 0 & 0 & 0 & 0 & 1 \\
				0 & 0 & 0 & 0 & 1 & 1 \\
				0 & 0 & 1 & 1 & 1 & 1 \\
				0 & 1 & 1 & 1 & 1 & 1 \\
				1 & 1 & 1 & 1 & 1 & 1 \\
				1 & 1 & 1 & 1 & 1 & 1 \\
				1 & 1 & 1 & 1 & 1 & 1
				\end{pmatrix}$ &
				$\begin{pmatrix} 0 & 0 & 0 & 0 & 0 & 0 \\
				0 & 0 & 0 & 0 & 0 & 0 \\
				0 & 0 & 0 & 0 & 0 & 0 \\
				0 & 0 & 0 & 0 & 0 & 1 \\
				0 & 0 & 0 & 0 & 1 & 1 \\
				0 & 0 & 1 & 1 & 1 & 1 \\
				1 & 1 & 1 & 1 & 1 & 1 \\
				1 & 1 & 1 & 1 & 1 & 1 \\
				1 & 1 & 1 & 1 & 1 & 1 \\
				1 & 1 & 1 & 1 & 1 & 1
				\end{pmatrix}$ & 
				$\begin{pmatrix} 0 & 0 & 0 & 0 & 0 & 0 \\
				0 & 0 & 0 & 0 & 0 & 0 \\
				0 & 0 & 0 & 0 & 0 & 0 \\
				0 & 0 & 0 & 0 & 0 & 0 \\
				0 & 0 & 0 & 0 & 1 & 1 \\
				0 & 0 & 1 & 1 & 1 & 1 \\
				1 & 1 & 1 & 1 & 1 & 1 \\
				1 & 1 & 1 & 1 & 1 & 1 \\
				1 & 1 & 1 & 1 & 1 & 1 \\
				1 & 1 & 1 & 1 & 1 & 1
				\end{pmatrix}$ & \\
				\bottomrule
			\end{tabular}
		\end{center}
		\caption{The optimal designs from Figure~\ref{fig1} are presented.}\label{tab7}
	\end{table}
	
	\section{D- and A-optimal designs from Section 3.5}\label{appopt}
	
	Here, in Tables~\ref{tabD} and \ref{tabA}, we provide the D- and A-optimal designs discussed in Section~\ref{across}.
	
	\begin{sidewaystable}[htbp]
		\renewcommand{\arraystretch}{1.2}
		\begin{center}
			\begin{tabular}{rcccccc}
				\toprule
				Factor && \multicolumn{5}{c}{D-optimal designs} \\
				\cline{1-1}\cline{3-7}
				$E(\rho)$ && $\{0,\dots,0.10\}$ & $\{0.11,\dots,0.19\}$ & $\{0.20,\dots,0.46\}$ & $\{0.47,0.48\}$ & 0.49 \\
				$X$ && $\begin{pmatrix} 0 & 0 & 0 & 0 & 0 & 0 \\
				0 & 0 & 0 & 0 & 0 & 0 \\
				1 & 1 & 1 & 1 & 1 & 1 \\
				1 & 1 & 1 & 1 & 1 & 1 \\
				2 & 2 & 2 & 2 & 2 & 2 \\
				2 & 2 & 2 & 2 & 2 & 2 \end{pmatrix}$ & $\begin{pmatrix} 0 & 0 & 0 & 0 & 0 & 0 \\
				0 & 0 & 0 & 0 & 0 & 1 \\
				0 & 1 & 1 & 1 & 1 & 1 \\
				1 & 1 & 1 & 1 & 1 & 2 \\
				1 & 2 & 2 & 2 & 2 & 2 \\
				2 & 2 & 2 & 2 & 2 & 2 \end{pmatrix}$ & $\begin{pmatrix}
				0 & 0 & 0 & 0 & 0 & 0 \\ 
				0 & 0 & 0 & 0 & 1 & 1 \\
				0 & 0 & 1 & 1 & 1 & 1 \\
				1 & 1 & 1 & 1 & 2 & 2 \\
				1 & 1 & 2 & 2 & 2 & 2 \\
				2 & 2 & 2 & 2 & 2 & 2 \end{pmatrix}$ & $\begin{pmatrix} 0 & 0 & 0 & 0 & 0 & 1 \\
				0 & 0 & 0 & 0 & 1 & 1 \\
				0 & 0 & 1 & 1 & 1 & 2 \\
				1 & 1 & 1 & 2 & 2 & 2 \\
				1 & 1 & 2 & 2 & 2 & 2 \\
				2 & 2 & 2 & 2 & 2 & 2 \end{pmatrix}$ & $\begin{pmatrix} 0 & 0 & 0 & 0 & 0 & 0 \\
				0 & 0 & 0 & 0 & 1 & 1 \\
				0 & 0 & 0 & 1 & 1 & 1 \\
				0 & 1 & 1 & 1 & 2 & 2 \\
				1 & 1 & 2 & 2 & 2 & 2 \\
				1 & 2 & 2 & 2 & 2 & 2 \end{pmatrix}$ \\
				\midrule
				$E(\rho)$ && $\{0.50,\dots,0.53\}$ & $\{0.54,0.55,0.56\}$ & 0.57 & $\{0.58,\dots,0.79\}$ & 0.80 \\
				$X$ && $\begin{pmatrix} 0 & 0 & 0 & 0 & 0 & 1 \\
				0 & 0 & 0 & 0 & 1 & 1 \\
				0 & 0 & 1 & 1 & 1 & 2 \\
				1 & 1 & 1 & 2 & 2 & 2 \\
				1 & 1 & 2 & 2 & 2 & 2 \\
				2 & 2 & 2 & 2 & 2 & 2 \end{pmatrix}$ & $\begin{pmatrix} 0 & 0 & 0 & 0 & 0 & 0 \\
				0 & 0 & 0 & 0 & 1 & 1 \\
				0 & 0 & 0 & 1 & 1 & 1 \\
				0 & 1 & 1 & 1 & 2 & 2 \\
				1 & 1 & 2 & 2 & 2 & 2 \\
				1 & 2 & 2 & 2 & 2 & 2 \end{pmatrix}$ & $\begin{pmatrix} 0 & 0 & 0 & 0 & 0 & 1 \\
				0 & 0 & 0 & 0 & 1 & 1 \\
				0 & 0 & 1 & 1 & 1 & 2 \\
				1 & 1 & 1 & 2 & 2 & 2 \\
				1 & 1 & 2 & 2 & 2 & 2 \\
				2 & 2 & 2 & 2 & 2 & 2 \end{pmatrix}$ & $\begin{pmatrix} 0 & 0 & 0 & 0 & 0 & 1 \\
				0 & 0 & 0 & 0 & 1 & 1 \\
				0 & 0 & 0 & 1 & 1 & 2 \\
				0 & 1 & 1 & 2 & 2 & 2 \\
				1 & 1 & 2 & 2 & 2 & 2 \\
				1 & 2 & 2 & 2 & 2 & 2 \end{pmatrix}$ & $\begin{pmatrix} 0 & 0 & 0 & 0 & 0 & 1 \\
				0 & 0 & 0 & 0 & 1 & 1 \\
				0 & 0 & 0 & 1 & 1 & 2 \\
				0 & 0 & 1 & 2 & 2 & 2 \\
				1 & 1 & 2 & 2 & 2 & 2 \\
				1 & 2 & 2 & 2 & 2 & 2 \end{pmatrix}$ \\
				\midrule
				$E(\rho)$ && 0.81 & 0.82 & 0.83 & 0.84 & 0.85 \\
				$X$ && $\begin{pmatrix} 0 & 0 & 0 & 0 & 0 & 1 \\
				0 & 0 & 0 & 0 & 1 & 1 \\
				0 & 0 & 0 & 1 & 2 & 2 \\
				0 & 1 & 1 & 2 & 2 & 2 \\
				1 & 1 & 2 & 2 & 2 & 2 \\
				1 & 2 & 2 & 2 & 2 & 2 \end{pmatrix}$ & $\begin{pmatrix} 0 & 0 & 0 & 0 & 0 & 1 \\
				0 & 0 & 0 & 0 & 1 & 1 \\
				0 & 0 & 0 & 1 & 1 & 2 \\
				0 & 0 & 1 & 2 & 2 & 2 \\
				1 & 1 & 2 & 2 & 2 & 2 \\
				1 & 2 & 2 & 2 & 2 & 2 \end{pmatrix}$ & $\begin{pmatrix} 0 & 0 & 0 & 0 & 0 & 1 \\
				0 & 0 & 0 & 0 & 1 & 1 \\
				0 & 0 & 0 & 1 & 2 & 2 \\
				0 & 1 & 1 & 2 & 2 & 2 \\
				1 & 1 & 2 & 2 & 2 & 2 \\
				1 & 2 & 2 & 2 & 2 & 2 \end{pmatrix}$ & $\begin{pmatrix} 0 & 0 & 0 & 0 & 0 & 1 \\
				0 & 0 & 0 & 0 & 1 & 1 \\
				0 & 0 & 0 & 1 & 1 & 2 \\
				0 & 0 & 1 & 2 & 2 & 2 \\
				1 & 1 & 2 & 2 & 2 & 2 \\
				1 & 2 & 2 & 2 & 2 & 2 \end{pmatrix}$ & $\begin{pmatrix} 0 & 0 & 0 & 0 & 0 & 1 \\
				0 & 0 & 0 & 0 & 1 & 1 \\
				0 & 0 & 0 & 1 & 2 & 2 \\
				0 & 0 & 1 & 2 & 2 & 2 \\
				1 & 1 & 2 & 2 & 2 & 2 \\
				1 & 2 & 2 & 2 & 2 & 2 \end{pmatrix}$ \\
				\midrule
				$E(\rho)$ && 0.86 & $\{0.87,0.88,0.89\}$ & 0.90 & $\{0.91,\dots,0.99\}$ & 1.00  \\
				$X$ && $\begin{pmatrix} 0 & 0 & 0 & 0 & 0 & 1 \\
				0 & 0 & 0 & 0 & 1 & 1 \\
				0 & 0 & 0 & 1 & 1 & 2 \\
				0 & 0 & 1 & 2 & 2 & 2 \\
				0 & 1 & 2 & 2 & 2 & 2 \\
				1 & 2 & 2 & 2 & 2 & 2 \end{pmatrix}$ & $\begin{pmatrix} 0 & 0 & 0 & 0 & 0 & 1 \\
				0 & 0 & 0 & 0 & 1 & 2 \\
				0 & 0 & 0 & 1 & 2 & 2 \\
				0 & 1 & 1 & 2 & 2 & 2 \\
				1 & 1 & 2 & 2 & 2 & 2 \\
				1 & 2 & 2 & 2 & 2 & 2 \end{pmatrix}$ & $\begin{pmatrix} 0 & 0 & 0 & 0 & 0 & 1 \\
				0 & 0 & 0 & 0 & 1 & 1 \\
				0 & 0 & 0 & 1 & 2 & 2 \\
				0 & 0 & 1 & 2 & 2 & 2 \\
				0 & 1 & 2 & 2 & 2 & 2 \\
				1 & 2 & 2 & 2 & 2 & 2 \end{pmatrix}$ & $\begin{pmatrix} 0 & 0 & 0 & 0 & 0 & 1 \\
				0 & 0 & 0 & 0 & 1 & 2 \\
				0 & 0 & 0 & 1 & 2 & 2 \\
				0 & 0 & 1 & 2 & 2 & 2 \\
				0 & 1 & 2 & 2 & 2 & 2 \\
				1 & 2 & 2 & 2 & 2 & 2 \end{pmatrix}$ & $\begin{pmatrix} 0 & 0 & 0 & 0 & 0 & 0 \\
				0 & 0 & 0 & 1 & 2 & 2 \\
				0 & 0 & 0 & 1 & 2 & 2 \\
				0 & 0 & 0 & 1 & 2 & 2 \\
				0 & 0 & 0 & 1 & 2 & 2 \\
				0 & 0 & 0 & 1 & 2 & 2 \end{pmatrix}$ \\
				\bottomrule
			\end{tabular}
		\end{center}
		\caption{D-optimal allocation matrices for cross-sectional designs with $D=3$. The D-optimal allocation matrices in the case $\mathfrak{I}=\{6\}$, $\mathfrak{C}=\{\mathfrak{C}_6\}=\{6\}$, $\mathfrak{M}=\mathfrak{M}_{6,6}=\{8\}$, and $\sigma^2=1$, with $w=0$ and $\beta=1$ are shown for $E(\rho)\in\{0,0.01,\dots,1\}$. No restrictions are placed on $\mathfrak{X}$ other than the identifiability of Equation~\ref{eq1}. Each allocation matrix was identified via our exhaustive search method.}\label{tabD}
	\end{sidewaystable}
	
	\begin{sidewaystable}[htbp]
		\renewcommand{\arraystretch}{1.2}
		\begin{center}
			\begin{footnotesize}
			\begin{tabular}{rcccccc}
				\toprule
				Factor && \multicolumn{5}{c}{A-optimal designs} \\
				\cline{1-1}\cline{3-7}
				$E(\rho)$ && $\{0,\dots,0.06\}$ & 0.07 & $\{0.08,\dots,0.11\}$ & $\{0.12,\dots,0.30\}$ & $\{0.31,0.32,0.33\}$ \\
				$X$ && $\begin{pmatrix} 0 & 0 & 0 & 0 & 0 & 0 \\
				0 & 0 & 0 & 0 & 0 & 1 \\
				1 & 1 & 1 & 1 & 1 & 1 \\
				1 & 1 & 1 & 1 & 1 & 1 \\
				1 & 2 & 2 & 2 & 2 & 2 \\
				2 & 2 & 2 & 2 & 2 & 2
				\end{pmatrix}$ & $\begin{pmatrix} 0 & 0 & 0 & 0 & 0 & 0 \\
				0 & 0 & 0 & 0 & 1 & 1 \\
				1 & 1 & 1 & 1 & 1 & 1 \\
				1 & 1 & 1 & 1 & 1 & 1 \\
				1 & 1 & 2 & 2 & 2 & 2 \\
				2 & 2 & 2 & 2 & 2 & 2
				\end{pmatrix}$ & $\begin{pmatrix} 0 & 0 & 0 & 0 & 0 & 0 \\
				0 & 0 & 0 & 0 & 1 & 1 \\
				0 & 1 & 1 & 1 & 1 & 1 \\
				1 & 1 & 1 & 1 & 1 & 2 \\
				1 & 1 & 2 & 2 & 2 & 2 \\
				2 & 2 & 2 & 2 & 2 & 2
				\end{pmatrix}$ & $\begin{pmatrix} 0 & 0 & 0 & 0 & 0 & 0 \\
				0 & 0 & 0 & 0 & 1 & 1 \\
				0 & 0 & 1 & 1 & 1 & 1 \\
				1 & 1 & 1 & 1 & 2 & 2 \\
				1 & 1 & 2 & 2 & 2 & 2 \\
				2 & 2 & 2 & 2 & 2 & 2
				\end{pmatrix}$ & $\begin{pmatrix} 0 & 0 & 0 & 0 & 0 & 0 \\
				0 & 0 & 0 & 0 & 1 & 1 \\
				0 & 0 & 1 & 1 & 1 & 1 \\
				1 & 1 & 1 & 1 & 2 & 2 \\
				1 & 1 & 2 & 2 & 2 & 2 \\
				1 & 2 & 2 & 2 & 2 & 2
				\end{pmatrix}$ \\
				\midrule
				$E(\rho)$ && $\{0.34,0.35\}$ & $\{0.36,\dots,0.39\}$ & $\{0.40,\dots,0.62\}$ & $\{0.63,\dots,0.66\}$ & 0.67 \\
				$X$ && $\begin{pmatrix} 0 & 0 & 0 & 0 & 0 & 0 \\
				0 & 0 & 0 & 0 & 1 & 1 \\
				0 & 0 & 1 & 1 & 1 & 1 \\
				0 & 1 & 1 & 1 & 2 & 2 \\
				1 & 1 & 2 & 2 & 2 & 2 \\
				1 & 2 & 2 & 2 & 2 & 2
				\end{pmatrix}$ & $\begin{pmatrix} 0 & 0 & 0 & 0 & 0 & 1 \\
				0 & 0 & 0 & 0 & 1 & 1 \\
				0 & 0 & 1 & 1 & 1 & 2 \\
				1 & 1 & 1 & 1 & 2 & 2 \\
				1 & 1 & 2 & 2 & 2 & 2 \\
				2 & 2 & 2 & 2 & 2 & 2
				\end{pmatrix}$ & $\begin{pmatrix} 0 & 0 & 0 & 0 & 0 & 1 \\
				0 & 0 & 0 & 0 & 1 & 1 \\
				0 & 0 & 1 & 1 & 1 & 2 \\
				0 & 1 & 1 & 1 & 2 & 2 \\
				1 & 1 & 2 & 2 & 2 & 2 \\
				1 & 2 & 2 & 2 & 2 & 2
				\end{pmatrix}$ & $\begin{pmatrix} 0 & 0 & 0 & 0 & 0 & 1 \\
				0 & 0 & 0 & 0 & 1 & 1 \\
				0 & 0 & 0 & 1 & 1 & 2 \\
				0 & 1 & 1 & 1 & 2 & 2 \\
				1 & 1 & 2 & 2 & 2 & 2 \\
				1 & 2 & 2 & 2 & 2 & 2
				\end{pmatrix}$ & $\begin{pmatrix} 0 & 0 & 0 & 0 & 0 & 1 \\
				0 & 0 & 0 & 0 & 1 & 1 \\
				0 & 0 & 1 & 1 & 1 & 2 \\
				0 & 1 & 1 & 2 & 2 & 2 \\
				1 & 1 & 2 & 2 & 2 & 2 \\
				1 & 2 & 2 & 2 & 2 & 2
				\end{pmatrix}$ \\
				\midrule
				$E(\rho)$ && $\{0.68,0.69\}$ & $\{0.70,\dots,0.81\}$ & 0.82 & 0.83 & $\{0.84,0.85,0.86\}$ \\
				$X$ && $\begin{pmatrix} 0 & 0 & 0 & 0 & 0 & 1 \\
				0 & 0 & 0 & 0 & 1 & 1 \\
				0 & 0 & 0 & 1 & 1 & 2 \\
				0 & 1 & 1 & 1 & 2 & 2 \\
				1 & 1 & 2 & 2 & 2 & 2 \\
				1 & 2 & 2 & 2 & 2 & 2
				\end{pmatrix}$ & $\begin{pmatrix} 0 & 0 & 0 & 0 & 0 & 1 \\
				0 & 0 & 0 & 0 & 1 & 1 \\
				0 & 0 & 0 & 1 & 1 & 2 \\
				0 & 1 & 1 & 2 & 2 & 2 \\
				1 & 1 & 2 & 2 & 2 & 2 \\
				1 & 2 & 2 & 2 & 2 & 2
				\end{pmatrix}$ & $\begin{pmatrix} 0 & 0 & 0 & 0 & 0 & 1 \\
				0 & 0 & 0 & 0 & 1 & 1 \\
				0 & 0 & 0 & 1 & 1 & 2 \\
				0 & 0 & 1 & 2 & 2 & 2 \\
				1 & 1 & 2 & 2 & 2 & 2 \\
				1 & 2 & 2 & 2 & 2 & 2
				\end{pmatrix}$ & $\begin{pmatrix} 0 & 0 & 0 & 0 & 0 & 1 \\
				0 & 0 & 0 & 0 & 1 & 1 \\
				0 & 0 & 0 & 1 & 2 & 2 \\
				0 & 1 & 1 & 2 & 2 & 2 \\
				1 & 1 & 2 & 2 & 2 & 2 \\
				1 & 2 & 2 & 2 & 2 & 2
				\end{pmatrix}$ & $\begin{pmatrix} 0 & 0 & 0 & 0 & 0 & 1 \\
				0 & 0 & 0 & 0 & 1 & 1 \\
				0 & 0 & 0 & 1 & 1 & 2 \\
				0 & 0 & 1 & 2 & 2 & 2 \\
				1 & 1 & 2 & 2 & 2 & 2 \\
				1 & 2 & 2 & 2 & 2 & 2
				\end{pmatrix}$ \\
				\midrule
				$E(\rho)$ && $\{0.87,0.88,0.89\}$ & $\{0.90,0.91,0.92\}$ & 0.93 & 0.94 & $\{0.95,0.96\}$ \\
				$X$ && $\begin{pmatrix} 0 & 0 & 0 & 0 & 0 & 1 \\
				0 & 0 & 0 & 0 & 1 & 2 \\
				0 & 0 & 1 & 1 & 2 & 2 \\
				0 & 1 & 1 & 2 & 2 & 2 \\
				1 & 1 & 2 & 2 & 2 & 2 \\
				1 & 2 & 2 & 2 & 2 & 2
				\end{pmatrix}$ & $\begin{pmatrix} 0 & 0 & 0 & 0 & 0 & 1 \\
				0 & 0 & 0 & 0 & 1 & 1 \\
				0 & 0 & 0 & 1 & 1 & 2 \\
				0 & 0 & 1 & 2 & 2 & 2 \\
				0 & 1 & 2 & 2 & 2 & 2 \\
				1 & 2 & 2 & 2 & 2 & 2
				\end{pmatrix}$ & $\begin{pmatrix} 0 & 0 & 0 & 0 & 0 & 1 \\
				0 & 0 & 0 & 0 & 1 & 2 \\
				0 & 0 & 0 & 1 & 2 & 2 \\
				0 & 1 & 1 & 2 & 2 & 2 \\
				1 & 1 & 2 & 2 & 2 & 2 \\
				1 & 2 & 2 & 2 & 2 & 2
				\end{pmatrix}$ & $\begin{pmatrix} 0 & 0 & 0 & 0 & 0 & 1 \\
				0 & 0 & 0 & 0 & 1 & 1 \\
				0 & 0 & 0 & 1 & 1 & 2 \\
				0 & 0 & 1 & 2 & 2 & 2 \\
				0 & 1 & 2 & 2 & 2 & 2 \\
				1 & 2 & 2 & 2 & 2 & 2
				\end{pmatrix}$ & $\begin{pmatrix} 0 & 0 & 0 & 0 & 0 & 1 \\
				0 & 0 & 0 & 0 & 1 & 1 \\
				0 & 0 & 0 & 1 & 2 & 2 \\
				0 & 0 & 1 & 2 & 2 & 2 \\
				0 & 1 & 2 & 2 & 2 & 2 \\
				1 & 2 & 2 & 2 & 2 & 2
				\end{pmatrix}$ \\
				\midrule
				$E(\rho)$ && 0.97 & $\{0.98,0.99\}$ & 1.00 \\
				$X$ && $\begin{pmatrix} 0 & 0 & 0 & 0 & 0 & 1 \\
				0 & 0 & 0 & 0 & 1 & 1 \\
				0 & 0 & 0 & 2 & 2 & 2 \\
				0 & 0 & 1 & 1 & 1 & 2 \\
				0 & 1 & 2 & 2 & 2 & 2 \\
				1 & 1 & 1 & 1 & 2 & 2
				\end{pmatrix}$ & $\begin{pmatrix} 0 & 0 & 0 & 0 & 0 & 1 \\
				0 & 0 & 0 & 0 & 1 & 2 \\
				0 & 0 & 0 & 1 & 2 & 2 \\
				0 & 0 & 1 & 2 & 2 & 2 \\
				0 & 1 & 2 & 2 & 2 & 2 \\
				1 & 2 & 2 & 2 & 2 & 2
				\end{pmatrix}$ & $\begin{pmatrix} 0 & 0 & 0 & 0 & 0 & 0 \\
				0 & 0 & 0 & 0 & 0 & 0 \\
				0 & 0 & 0 & 0 & 0 & 0 \\
				1 & 2 & 2 & 2 & 2 & 2 \\
				1 & 2 & 2 & 2 & 2 & 2 \\
				1 & 2 & 2 & 2 & 2 & 2
				\end{pmatrix}$ \\
				\bottomrule
			\end{tabular}
			\end{footnotesize}
		\end{center}
		\caption{A-optimal allocation matrices for cross-sectional designs with $D=3$. The A-optimal allocation matrices in the case $\mathfrak{I}=\{6\}$, $\mathfrak{C}=\{\mathfrak{C}_6\}=\{6\}$, $\mathfrak{M}=\mathfrak{M}_{6,6}=\{8\}$, and $\sigma^2=1$, with $w=0$ and $\beta=1$ are shown for $E(\rho)\in\{0,0.01,\dots,1\}$. No restrictions are placed on $\mathfrak{X}$ other than the identifiability of Equation~\ref{eq1}. Each allocation matrix was identified via our exhaustive search method.}\label{tabA}
	\end{sidewaystable}
	
	\section{Application to binary outcome variables}\label{app2}
	
	In this section, we provide a brief description of how our methods can be applied to binary outcome variables (in the case $D=2$ for a cross-sectional design). Analysing at the cluster level, the following hierarchical model can be utilised for data analysis
	\begin{align*}
	r_{ij} &\sim Bin(m,p_{ij}),\\
	\text{logit}(p_{ij}) &= \mu + \pi_j + \beta_{1}X_{ij} + c_i + \epsilon_{ij},
	\end{align*}
	where $c_i\sim N(0,\sigma_c^2)$ and $\epsilon_{ij}\sim N(0,\sigma_\epsilon^2/m)$. Moreover, $r_{ij}$ is the number of responses observed in cluster $i$ in period $j$, and $p_{ij}$ is therefore the probability of response in cluster $i$ in period $j$.
	
	We can then apply our methodology by assuming that $\sigma_e^2/m=1/\{m\bar{p}(1 - \bar{p})\}$, where $\bar{p}$ is the average response rate. In practice, one would need to then assess the performance of the approximation via a simulation study to assess the empirical power of identified efficient designs. As discussed in Section~\ref{disc}, we may reasonably anticipate that such approximation based results are likely to only be reliable for large sample sizes.

\clearpage
\nocite{*}
\bibliographystyle{abbrvnat}
\bibliography{arxiv}

\end{document}